\documentstyle[12pt]{article}
\makeatletter
\newdimen\normalarrayskip              
\newdimen\minarrayskip                 
\normalarrayskip\baselineskip
\minarrayskip\jot
\newif\ifold             \oldtrue            \def\new{\oldfalse}
\def\arraymode{\ifold\relax\else\displaystyle\fi} 
\def\eqnumphantom{\phantom{(\theequation)}}     
\def\@arrayskip{\ifold\baselineskip\z@\lineskip\z@
     \else
     \baselineskip\minarrayskip\lineskip2\minarrayskip\fi}
\def\@arrayclassz{\ifcase \@lastchclass \@acolampacol \or
\@ampacol \or \or \or \@addamp \or
   \@acolampacol \or \@firstampfalse \@acol \fi
\edef\@preamble{\@preamble
  \ifcase \@chnum
     \hfil$\relax\arraymode\@sharp$\hfil
     \or $\relax\arraymode\@sharp$\hfil
     \or \hfil$\relax\arraymode\@sharp$\fi}}
\def\@array[#1]#2{\setbox\@arstrutbox=\hbox{\vrule
     height\arraystretch \ht\strutbox
     depth\arraystretch \dp\strutbox
     width\z@}\@mkpream{#2}\edef\@preamble{\halign \noexpand\@halignto
\bgroup \tabskip\z@ \@arstrut \@preamble \tabskip\z@ \cr}%
\let\@startpbox\@@startpbox \let\@endpbox\@@endpbox
  \if #1t\vtop \else \if#1b\vbox \else \vcenter \fi\fi
  \bgroup \let\par\relax
  \let\@sharp##\let\protect\relax
  \@arrayskip\@preamble}
\def\eqnarray{\stepcounter{equation}%
              \let\@currentlabel=\theequation
              \global\@eqnswtrue
              \global\@eqcnt\z@
              \tabskip\@centering
              \let\\=\@eqncr
              $$%
 \halign to \displaywidth\bgroup
    \eqnumphantom\@eqnsel\hskip\@centering
    $\displaystyle \tabskip\z@ {##}$%
    &\global\@eqcnt\@ne \hskip 2\arraycolsep
         \hfil$\arraymode{##}$\hfil
    &\global\@eqcnt\tw@ \hskip 2\arraycolsep
         $\displaystyle\tabskip\z@{##}$\hfil
         \tabskip\@centering
    &{##}\tabskip\z@\cr}
\makeatother
\def\Bf#1{\mbox{\boldmath $#1$}}

\def\beq{\begin{equation}}
\def\eeq{\end{equation}}
\def\bea{\begin{eqnarray}}
\def\eea{\end{eqnarray}}

\def\dst{\displaystyle}

\begin{document}

\begin{titlepage}
\begin{center}
{{\it P.N.Lebedev Institute preprint} \hfill FIAN/TD-02/92\\
{\it I.E.Tamm Theory Department} \hfill ITEP-M-2/92
\begin{flushright}{February 1992}\end{flushright}
\vspace{0.1in}{\Large \bf $c  =  r_G$ Theories of $W_G$-gravity:}\\{\Large
\bf the set of observables}\\{\Large \bf as a {\it model} of simply laced
$G$}\\[.8in]
{\large  A.Marshakov, A.Mironov}\\
\bigskip {\it Theory Department,\\P.N.Lebedev Physical
Institute,\\ Leninsky prospect, 53, Moscow, 117 924},
\footnote{E-mail address: theordep@sci.fian.msk.su}\\ \smallskip
\bigskip {\large A.Morozov\footnote{E-mail address: morozov@itep.msk.su},
M.Olshanetsky\footnote{E-mail address: olshanetsky@itep.msk.su}}\\
 \bigskip {\it Institute of Theoretical and Experimental
Physics,  \\
 Bol.Cheremushkinskaya st., 25, Moscow, 117 259}}
\end{center}
\bigskip
\bigskip

\newpage
\setcounter{page}2
\centerline{\bf ABSTRACT}
\begin{quotation}

We propose to study a generalization of the Klebanov-Polyakov-Witten (KPW)
construction for the algebra of observables in the  $c = 1$  string model to
theories with  $c > 1$. We emphasize the algebraic meaning of the KPW
construction for  $c = 1$  related to occurrence of a {\it model} of
{\it SU}(2) as original structure on the algebra of observables. The attempts
to preserve this structure in generalizations naturally leads to consideration
of $W$-gravities. As a first step in the study of these generalized KPW
constructions we design explicitly the subsector of the space of observables in
appropriate  $W_G$-string theory, which forms the {\it model} of $G$ for any
simply laced {\it G}. The {\it model} structure is confirmed by the fact that
corresponding one-loop Kac-Rocha-Caridi $W_G$-characters for  $c = r_G$ sum
into a chiral (open string) $k=1$ $G$-WZW partition function.
\end{quotation}
\end{titlepage}

\newpage
\setcounter{footnote}0
\section{Introduction}

Two recent papers [1,2] began a systematic study of the what can be called the
{\it algebras of observables} in $c = 1$ theories of $2d$-gravity. These
algebras
should be considered as main invariant characteristics of topological (and/or
string) models (in particular they should not depend on the order of
perturbation expansion). In most cases string models are constructed starting
from particular conformal field theory (CFT) by coupling to and integration
over 2-dimensional metric on Riemann surfaces. In practice, one usually
proceeds with a kind of BRST formalism instead of explicit integration:
({\it i})
extra Liouville and ghost fields are added to degrees of freedom of original
CFT in order to create a new conformal model with vanishing Virasoro central
charge and (ii) the action of Virasoro algebra is factorized
out\footnote{Really
these two items are related: when the central charge is equal to
zero one can pull out Virasoro action from descendants to the vacuum and
cancell corresponding correlators, then only correlators of primaries are
non-zero.},
$i.e.$ only the classes of BRST cohomology are identified with the
observables in a string model. Moreover, at least, in many cases certain
Virasoro-primary fields of original CFT, appropriately ``dressed" by Liouville
and ghost fields, can be taken for representatives of the cohomology classes.
This is a considerable simplification, because the set of primaries can possess
additional structures or symmetries which are much more obscure in the entire
huge set of all fields of CFT. As an example of such structure, refs.[1,2]
associated the set of all Virasoro primaries $SU(2)$-invariant
$c=1$ CFT to the {\it model} of $SU(2)$, $i.e.$ to the collection of all
the unitary highest-weight representations of $SU(2)$ where each
representation appears once and only once. This {\it model} structure easily
survives Liouville dressing and characterizes (a certain subset of) the set of
observables in the corresponding string model (most transparently in the chiral
or open-string sector). Moreover, it predetermines, to some extent, the form of
the algebra of observables in the latter theory.

The main logical steps in the scheme of [1,2] may be described as follows:

\bigskip

{\bf 1)}
Consider first the matter (CFT) sector. Take one free scalar field  $X$,
compactified on a circle of radius  $r = \sqrt{2}$ ($i.e.$ $X \sim
X + 2\pi r =
X +2\pi \sqrt{2})$, with a stress tensor
$\displaystyle{-{1\over 2}(\partial X)^2}$. At this
value of radius the original symmetry $U(1)\times U(1)$ enlarges to
$SU(2)\times SU(2)$, therefore one can consider chiral sector with symmetry
$SU(2)$). This correspond to the self-dual point of the  $c = 1$
Gaussian model [3,4] where $SU(2)\times SU(2)$ acts naturally on the Virasoro
primaries of the theory\footnote{Thus,
one can single out the self-dual radius as a point of enhanced
symmetry group whose {\it model} we will be trying to reproduce.}.
In what follows we would rather consider the holomorphic or open-string
sector of the theory (like in [1]) which can be naturally decomposed with
respect to single  $SU(2)$  action. Then all the primary ``chiral"
vertex operators in such matter theory form a {\it model} of $SU(2)$:
$M[SU(2)].$

{\it Comment:} Thus, the loop chiral algebra of this theory, considered as CFT,
${\cal J}_{\pm} = e^{\pm \imath \sqrt{2}x}$ and  ${\cal J}_{_0}=
i\sqrt{2}\partial x$
($x$  is the holomorphic part of the scalar field  $X$). Then the generators
$\displaystyle{Q_\pm  =
\oint   {\cal J}_\pm }$,  $\displaystyle{Q_{_0}= \oint   {\cal J}_{_0}}$
form usual
$SU(2)$ algebra\footnote{Since
$X \sim  X+2\pi \sqrt{2}$,  $Q_{_0}$ does not need to vanish.}.
The highest-weight operators ($i.e.$ those annihilated by $Q_+$) are given by
$\psi _{J,J} = e^{i\sqrt{2}Jx}$ with any (half-)integer  {\it J}. Since
$SU(2)$  generators commute with the stress tensor, their action
transform Virasoro primaries into primaries, and whole sequence  $\psi _{J,m}
\equiv  :Q_-\psi _{J,m+1}:$  beginning from  $\psi _{J,J}$ consists of primary
fields. For any (half-)integer $J$ this sequence is finite: $\psi _{J,-J} =
e^{-i\sqrt{2}Jx}$ is the lowest-weight operator ($i.e.$ is annihilated by
$Q_-$). The set $\{\psi _{J,m}\}$ with $|m|\leq J$ and given $J$ forms a spin
$J$ representation of $SU(2)$. All these operators  $\psi _{J,m}$ are
the Virasoro primaries and there are no other primaries\footnote{For
other compactification radii the set of primaries is different.}.
This last statement is not quite obvious since it is not easy to exclude any
other candidates for the role of highest weight fields as well as the possible
occurrence of non-highest-weight representation. The simplest
way\footnote{Alternative
(and long) way is, for example, to thoroughly investigate the
operator algebra.}
to confirm that no primaries are overlooked is to reproduce the holomorphic (or
open-string) partition function from a sum over all characters of irreducible
representations of the Virasoro algebra associated with the announced set of
primaries\footnote{An
alternative description of the same chiral theory is in terms of the
level  $k=1\ \ WZW$ model.}.

{\it Conclusion:} Thus at the starting point one already discovers a
{\it model} of $SU(2)$: it is formed by the set of primaries of
corresponding conformal theory. This set is, of course, not the same as the
entire spectrum of the matter conformal theory, since operator product
expansion (OPE) in CFT contain also descendants, which have nothing to do with
the structure of $M[SU(2)]$. In order to get a $M[SU(2)]$ structure one has to
get rid of all Virasoro-descendants.

\bigskip

{\bf 2)}
In order to achieve this goal let us couple the matter sector to the $2d$
gravity, $i.e$. consider the corresponding string model. There is a well
defined subsector in the space of observables, which consists of surface (for
holomorphic part contour-) integrals of ({\it matter}$)+(${\it Liouville})
primaries,
provided the full dimension is one.

{\it Comment}: The drastic reduction of the space of vertex operators is a
general phenomenon occurring whenever a string model is build from CFT. Let us
remind first the situation with the critical string. In the simplest treatment,
one neglects Liouville field at all, but this is not the only thing one does.
There are three additional requirements to physical vertex operators in
critical string models:

a) They do not contain ghost fields (in certain and natural
picture)\footnote{Except for one delicate point, concerning dilaton operator.}.

b) They are Virasoro primaries (with respect to ``full" Virasoro). (All the
descendants, present in CFT are ``gauged out". The reason is that coupling to
$2d$ gravity implies gauging the Virasoro algebra and thus all the descendants
are eliminated as gauge-non-invariant operators. They really decouple from
the correlators\footnote{It is just no-ghost theorem.}).

c) They are integrals (in the picture consistent with  a)) of operators of
conformal dimension one\footnote{Let
us note that other (equivalent) pictures may be obtained by
multiplication of dimension one integrands (constrained by a) and b)) with
ghost field instead of integration - see also sect.3.}.

Naive generalization of these three principles to the case of non-critical
string has been suggested in [5,6], with the only modification made at the
point b), where one is supposed to take primaries of the full ({\it matter} +
{\it Liouville}) Virasoro algebra, in the form of ({\it matter
primaries})$\times$({\it Liouville  primaries}) and
Liouville primaries being pure
exponentials. It is, however, not quite true that such a simple generalization
is valid. A problem arises at the very beginning point a). In order to
{\it derive} honestly this requirement one should prove that in every BRST
cohomology class there is a ghost free representative. According to [7,2] this
is not true in non-critical case: there are two additional representatives of
the cohomology classes, which unavoidably (and non-trivially) contain ghost
fields. In the particular case of $c=1$ model, the first ones were interpreted
in [2] as a ground ring, another one (the ``ghost number two") still has no
nice interpretation. It seems, however, that the subsector, described by
principles a),b),c) is closed by itself under OPE (modulo descendants, $i.e.$
fields vanishing in the correlation functions). Another delicate point is that
the requirement c), $i.e.$ that the full dimension is one, permits two
different choices of Liouville primaries associated with a given matter
operator. In what follows we consider a subsector with one specific choice of
these two. This subsector also seems to be closed under OPE in the above sense.

{\it Conclusion}: In any string model ($i.e.$ CFT coupled to $2d$ gravity) a
certain subset of observables is in one-to-one correspondence with the set of
primaries of the matter theory. Therefore, if we start from the matter model,
described in 1), the space of observables contains a subset, which is a
{\it model} of $SU(2)$: it is formed by the set of primaries of
$M[SU(2)]$  and is closed under OPE modulo descendants. The
corresponding vertex operators are of the form

\beq
q_{J,m} = \oint \psi _{J,m}(x)e^{(J-1)\sqrt{2}\phi }\ .
\eeq

\bigskip

{\bf 3)} These operators form a Lie algebra  ${\cal G}$  (in contrast to the
OPE in
original CFT, which is not a Lie algebra), its structure constants being
defined by 3-point correlation functions on a sphere.

{\it Conclusion:} Thus, starting from 1) we obtain an embedding of  $M[SU(2)]$
into a certain subset of algebra of observables, which turns to be a {\it Lie}
algebra  ${\cal G}[SU(2)]$. Moreover, according to 1) this embedding $M[SU(2)]
\hookrightarrow {\cal G}[SU(2)]$ is in fact a representation ($i.e.$ respects
the structure of a {\it model}):

a) $q_{1,m}= Q_m$, and thus form the {\it adjoint} representation of $SU(2)$
in ${\cal G}[SU(2)];$

b) since $Q_m$ do not act on the Liouville field $\phi $, $\{q_{J,m}\}$ form
{\it model} of $SU(2)$: it is formed by the set of primaries of
a spin  $J$ representation  generated by  $\{Q_m\}$, in
${\cal G}[SU(2)].$

Thus, Klebanov-Polyakov-Witten (KPW) construction gives rise to {\it model}
$M[SU(2)]$ as representation of the Lie algebra  ${\cal G}[SU(2)]$. This
algebra defines an additional structure: while on  $M[SU(2)]$  (which is a
priori a collection of representations of $SU(2)$  with every
representation appearing exactly once) only these commutators are defined
where, at least, one of the items is  $Q_m=q_{1,m}$, in  ${\cal G}[SU(2)]$ the
commutation of {\it any} two elements  $q_{J',m'}$ and $q_{J'',m''}$ make
sense.

\bigskip

{\bf 4)} Moreover,
due to specific selection rules (dictated by the properties of
Liouville sector)  $[q_{J',m'}$, $q_{J'',m''}]$  contain a single
representation  $q_{J,m}$ with  $J=J'+J''-1$  (and $m=m'+m''$):

\beq
[q_{J',m'}\hbox{, } q_{J'',m''}] = C^{J'+J''-1}_{J',J''}q_{J'+J''-1,m'+m''}\ .
\eeq
The coefficients  $C[SU(2)]$  are  $3j$-symbols (Clebsh-Gordon coefficients) of
$SU(2)$  in a certain basis.

\bigskip

{\bf 5)} This Lie algebra with structure constants defined by  $3j$-symbols
$C^{J'+J``-1}_{J',J``}$ may be alternatively interpreted as an algebra of
area-preserving diffeomorphisms of a 2-dimensional plane  ${\Bf R}^2 \sim
{\Bf C}$ ($i.e.$ as an algebra of Hamiltonian vector fields\footnote{See
Appendix.}
on a plane, which can be also identified with a realization of
$W_\infty $-algebra).

\bigskip

{\bf 6)} ${\cal G}[SU(2)]$
may be also identified with the algebra of derivatives of
the ground ring which is formed by the other piece of the algebra of
observables: of dimension zero and ghost number zero physical vertex operators
[2,7]. The ground ring is, in fact, isomorphic to the ring of Hamiltonians
(polynomials on ${\Bf R}^2 \sim  {\Bf C}$).

\bigskip
This was a brief description of the original KPW construction for the case of
$G=SU(2)$. If we omit all the details, the result is as follows:

{\it Consideration of OPE of physical operators (modulo fields vanishing in
physical correlators) in a certain  $c=1$  string model (1,2) suggests
the representation of a {\it model} $M[SU(2)]$ in a Lie algebra
${\cal G}[SU(2)]$
(3), with structure constants defined by certain $3j$-symbols of $SU(2)$
(4) and isomorphic to an algebra $W_\infty $ of area-preserving
diffeomorphisms of ${\Bf R}^2 \sim  {\Bf C}$ (5), while the ring of
Hamiltonians may be interpreted as the ground ring of $\dim =0,
\# ghost=0$ physical operators (6)}.

This is a more or less nice description of the algebra of observables in the
$SU(2)$-symmetric  $c = 1$  string model and can serve as a sample
example of what one should try to achieve when studying algebras of observables
in other string models. Some fragments of such description are already
available in  $c < 1$  case, where an essential piece of the algebra of
observables is given by the Virasoro- and $W$-constraints, which are the
appropriate counterparts (and, in fact, fragments) of  $W_\infty $ in the  $c =
1$  case.

Another appealing property of the $c=1$ example is its pure algebraic aspect:
the possibility to introduce a new algebraic structure in the {\it model} of a
simple Lie algebra $G=SU(2)$, which in this particular case is just a
Lie-algebra commutator with the structure constants given by appropriate
$3j$-symbols of original $G$, and, moreover, possessing an interpretation as an
algebra of diffeomorphisms of an orbit of  $G$.

It is very natural to look for generalizations of these results, which seem
interesting both from physical and mathematical points of view. What we propose
to do is to start from any simply laced algebra  $G$  and try to work out the
analogue of KPW construction as far as possible. This paper concerns with steps
1) and 2) of the above scheme, which are rather straightforward, but
unavoidably involve the concepts of  $W_G$-gravity and  $W_G$-strings, which
are still very poorly understood. We believe that their natural occurrence in
the KPW construction demonstrates once again that $W_G$-gravities are not
artificial ``game of the mind" in the context of string theory for $c>1$
and should
attract much more attention to the interesting field of $W$-geometry [8-13].

In section 2 we consider the spectrum of physical fields in the chiral sector
of  $c = r_G$ $W_G$-string theory. We prove in sect.2.2 that the fact that they
form a {\it model} of $G$ can be explicitely checked by the one-loop partition
function calculations in chiral (or open) theory. In other words, we prove that
the chiral (or open string) sector of $k=1$ $G$-WZW theory forms, in fact, a
{\it model} of compact Lie group {\it G}. In sect.2.3 we discuss the spectrum
of matter theory by BRST methods and in section 3 consider coupling to
$W_G$-gravity. Finally, the Appendix contains some facts about connection of
{\it model structure} with symplectomorphism algebras on various coadjoint
orbits of
$G$.

\section{Primary fields for compactifications on root lattices}

There are two different approaches to the problem -- we present both of them.
The first is technically simpler, the second -- more sophisticated, but
instead it emphasizes the important relation to the structure of Verma modules
and null-vectors. We consider these two approaches separately.

\subsection{Circle compactification (the case of  $SU(2)$)}

{\it The model structure.} Consider a theory of one free scalar field  $X$,
compactified on a circle, with Lagrangian  $\int   \partial X\bar \partial X$.
Its chiral algebra usually contains $\hat U(1)\times \hat U(1)$, generated by
${\cal J}_{_0}= \partial X$, and Virasoro algebra, generated by  $T = -
{1\over 2}(\partial X)^2$ (times their conjugate). The second one belongs to
the $\hat U(1)$ universal envelope. Normally the set of primary fields in this
theory is given by  $e^{ipx}e^{i\bar p\bar x}$ \footnote{$x$
and $\bar x$ denote holomorphic and anti-holomorphic parts of $X$
respectively.}, where

\beq
\new
\begin{array}{c}
p + \bar p = {n\over R}\ ,\\
p - \bar p = 2mR\ ,
\end{array}
\eeq
$n,m \in  {\Bf Z}$, where $R$ is so-called radius of compactification [3,4] (to
avoid misunderstanding we stress that it is really a {\it half} of real
compactification radius: $x \sim  x+2\pi R$  and  $\bar x \sim  \bar x+2\pi R$,
but  $x+\bar x = X \sim  X+2\pi r$,  where  $r = 2R)$. However, sometimes the
holomorphic chiral algebra is enlarged to become $S\hat U(2)_{k=1}$ generated
by ${\cal J}_\pm = e^{\pm i\sqrt{2}x}$ and  ${\cal J}_{_0}$. This happens in
the self-dual point when $R=1/\sqrt{2}$ (this theory is, of course, related to
the  $\hat SU(2)_{k=1}$ WZW model, see below). In this case the set of
primaries
should be $SU(2)\times SU(2)$-invariant, for holomorphic or open-string theory
it means that with any $e^{ipx}$ the spectrum should contain all non-vanishing
$(Q_-)^ke^{ipx}$ (if $p>0$, or $(Q_+)^ke^{ipx}$ if $p<0$; $Q_\pm = \oint
{\cal J}_\pm $ are generators of $SU(2)$ and commute with the stress
tensor). Whenever $p = integer\times \sqrt{2}$ this sequence is finite:
$k\leq |p|/\sqrt{2}$ and form a spin $J$ ($J=|p|/2)$ representation of
$SU(2)$. Clearly every representation appears exactly once and we
obtain a {\it model} of $SU(2)$.

\bigskip

{\it Holomorphic partition functions as a character of a {\it model}.} This
conclusion is confirmed by the formula for one-loop partition function of the
theory. Indeed, it equals [3,4,14] for $R = 1/\sqrt{2}$, ($q =
e^{2\pi i\tau }$)

\beq
{\cal Z}(\tau ,\bar \tau ) =
{\left| \theta \left[ {0}\atop {0}\right] (2\tau )\right| ^2 +
\left| \theta \left[ {1/2}\atop {0}\right] (2\tau )\right| ^2\over |\eta
(q)|^2}
\eeq
where  $\theta $ -- are ordinary Jacobi theta-functions, and coincides with the
partition function of  $SU(2)_{k=1}$ WZW theory. Let us introduce the quantity

\beq
Z(\tau ) = {\theta \left[ {0}\atop {0}\right] (2\tau )\over \eta (q)} +
{\theta \left[ {1/2}\atop {0}\right] (2\tau )\over \eta (q)}\hbox{ .}
\eeq
This can be interpreted as a ``holomorphic square root"\footnote{This
name may be to some extent motivated by the famous examples of
$E(8)$ and $SO(32)$ when the weight lattices are equivalent to the root
ones (even self dual lattices $\Gamma _8$ and $\Gamma _{16}$ respectively) and
there is only a single term in the sum (1).}
of partition function or, what is essentially the same, as partition functions
of the corresponding open-string models [15-17]. This holomorphic partition
function can be represented as a certain sum of the Virasoro characters over
all the Virasoro primaries, appearing in the spectrum of the theory. In
accordance with our arguments above (2) exactly the {\it model} of all
representations of $SU(2)$:

\beq
Z(\tau ) = \sum _{J\in {\Bf Z}_{+}/2} (2J+1)\chi _J(\tau )
\eeq
Here  $\chi _J(\tau )$  denote the Kac characters of representations Virasoro
algebra for  $c = 1$  [18]:

\beq
\chi _J(\tau ) = {q^{J^2}- q^{(J+1)^2}\over \eta (q)}.
\eeq
and multiplicities  $(2J+1) = \dim R_J$ reflect
that we indeed have the {\it model} of
$SU(2)$. Substitution of (7) into (6) gives:

\beq
\new
\begin{array}{c}
\eta (q)Z(\tau ) = {1\over 2}\left\lbrace  \sum _{J \in {\Bf Z}_{+}/2}
(2J+1)\chi _J(\tau )  +  (J \rightarrow  -J-1)\right\rbrace  =
\sum ^\infty _{n=-\infty }  q^{(n/2)^2} =\\
= \theta \left[ {0}\atop {0}\right] (\tau /2) = \theta \left[ {0}\atop {0}
\right] (2\tau ) +
\theta \left[ {1/2}\atop {0}\right] (2\tau ),
\end{array}
\eeq
in accordance with (5).

{\it KRC-formula from a limiting procedure.} As to the formula (7), it may be
easily derived\footnote{This
derivation of a character for a trivial case from a non-trivial one
is of course not the most clever thing to do, but these Rocha-Caridi characters
arise in the study of minimal models and thus their  $c < r_G$ analogues are
much better known than the analogues of  $c=1$ characters. What is even more
important, for $W$-algebras the counterparts of these ``minimal" Rocha-Caridi
characters are easily available [23], and it will be simpler for us below to
take their limits as $c$ approaches $r_G$, than perform an independent
derivation of $W_G$-characters for $c=r_G$.}
from the well-known expressions [19] for the Rocha-Caridi
characters of the Virasoro algebra for $\displaystyle{c = 1 -
{6(p-p')^2\over pp'} \equiv  1
- 12\alpha ^2_0} $,

\beq
\new
\begin{array}{c}
\chi _{m,n}(\tau ) = {q^{- {1\over 4pp'}}\over \eta (q)} [\theta \left[ {
{mp-np'}\over {2pp'}}\atop {0} \right] (0|2pp'\tau ) -  \theta \left[ {
{mp+np'}\over {2pp'}}\atop {0} \right] (0|2pp'\tau )] =\\
= {1\over \eta (q)}[q^{{1\over 4pp'}\{(mp-np')^2-1\}} -
q^{{1\over 4pp'}\{(mp+np')^2-1\}} +\ldots ],
\end{array}
\eeq
in the limit  $p' = p+1 \rightarrow  \infty $ (condition  $p'=p+1$  is imposed
in order to guarantee the unitarity\footnote{Really
we only need the difference $p'-p$ to be finite when $p
\rightarrow  \infty $.}).
Irreducible representations of Virasoro algebra are unambiguously labeled by
their dimensions and the character (5) corresponds to

\beq
\Delta _{m,n} = {(m\alpha _++ n\alpha _-)^2-(\alpha _++ \alpha _-)^2\over 8}\ .
\eeq
Here $\alpha _\pm  = \alpha _0 \pm  \sqrt{2+\alpha ^2_0}$, $\alpha _+ =
\sqrt{2p/p'}$, $\alpha _-= -\sqrt{2p'/p}$,  $\alpha _+\alpha _- = -2$,
$\alpha _++\alpha _-= 2\alpha _0$. In the limit $p'=p+1 \rightarrow  \infty $
these parameters turn into  $\alpha _\pm  \rightarrow  \pm \sqrt{2}$,
$\alpha _0 \rightarrow  0$  and  $\Delta _{m,n} \rightarrow
\tilde \Delta _{m,n} \equiv  \displaystyle{{(n-m)^2\over 4}}$.

The algebraic sum at the $r.h.s.$ of (5) arises because one needs to eliminate
the sub-modules from the Verma module, which are generated by null-vectors.
Such null-vectors arise at the levels  $n\cdot m$, $(p-n)\cdot (p'-m)$,
{\it etc}.  in the original module and are naturally decomposed into two
different sets. The levels from one set depend strongly on $p'$ and/or $p$ and
go to infinity when $p'=p+1 \rightarrow  \infty $. This leads to a drastic
simplification in the case of $c=1$: all intersections of sub-modules
disappear, the remaining sub-modules form a single ``tower" of embedded
modules, and instead of an infinite algebraic sum in (5) a single subtraction
survives:  $\chi _{m,n}(\tau ) \rightarrow  \tilde \chi _{m,n}(\tau ) \equiv
\displaystyle{{1\over \eta (q)}[q^{{(n-m)^2\over 4}} - q^{{(n+m)^2\over 4}}]}$.
This is,
however, not the whole story:  $\tilde \chi (\tau )$  does not need to be a
character of an {\it irreducible} representation at  $c=1$. This is explained
by appearance of new null-vectors in the limit $c=1$, which were absent for all
values of $c<1$ ($\alpha _0\neq 0$). This is indeed the case, unless either $m$
or $n$ is equal to 1. In fact, this is already clear from the fact, that all
dimensions of primaries  $\Delta _{m,n}$ (10) with given  $n-m$ coincide in the
limit $p'=p+1 \rightarrow  \infty $, and thus all such representations of the
Virasoro algebra should become identical. It means that representations with
$n$ (or $m$) $\neq  1$  should acquire new null-vectors when  $c$  becomes 1.

To make this statement a bit more formal, let us introduce a primary field

\beq
\Phi _{m,n}= \exp \left\{ \imath{1\over 2}[(m+1)\alpha _+ +
(n+1)\alpha _-]x\right\}
\eeq
associated with the character  $\chi _{m,n}$. Denote by  ${\cal L}_{-N} =
L_{-N} + \ldots$
the algebraic combination of Virasoro generators  $L_{-i}$ ($i>0$),
which can create the null-vector in the  $\Phi _{m,n}$ module at level $N$.
Then for any minimal model there exists a primary field at level  $N = mn = 2Jn
+ n^2$, but only for  $\alpha _0=0$  a {\it new} primary field appears also at
$N = m = 2J+1$  (here  $2J = m-n$); in terms of zero norms it means that
$\|{\cal L}_{-N}\Phi _{m,n}\|^2 = 0$  for $N = mn = 2Jn + n^2$, while
$\|{\cal L}_{-N}\Phi _{m,n}\|^2 \sim  \alpha ^2_0$ for  $N = 2J+1$. Thus, a new
primary field of this kind appears exactly in the limit  $\alpha _0
\rightarrow  0$, $i.e.$ when  $c \rightarrow  1$. For example, in the simplest
case of  $N=2$ the possible candidate to be the new primary field at the second
level should have the form

\beq
[a(\partial x)^2 + b\partial ^2x]e^{i\alpha x}\ .
\eeq
One has to require the OPE of (12) with the stress-tensor
$\displaystyle{-{1\over 2}(\partial x)^2+i\alpha _0\partial ^2x}$  being of the
usual form for
the primary field, in particular this means that third- and fourth-order poles
in the OPE vanish, what corresponds to

\beq
\left\{
\begin{array}{c}
-a + 2i(3\alpha _0 - \alpha )b = 0\\
i(2\alpha _0 - \alpha)a + b =0
\end{array}
\right.\ .
\eeq
The system (13) has non-zero solutions only with zero determinant, $i.e.$

\beq
6\alpha ^2_0 - 5\alpha _0\alpha  + \alpha ^2 = {1\over 2}
\eeq
with obvious solutions

\beq
\alpha  = {5\over 2}\alpha _0 \pm  {1\over 2}\sqrt{\alpha ^2_0+2}\ .
\eeq
Comparing (15) (for example for the positive sign) with the formula (11):

\beq
\alpha  = \alpha _{m,n} = {1\over 2}[(m+1)\alpha _++(n+1)\alpha _-] =
{m+n+2\over 2} \alpha _0 + {m-n\over 2}\sqrt{\alpha ^2_0+2}\ ,
\eeq
we get two distinguished cases: for  $\alpha _0\neq 0$, $m=2$, $n=1$ give the
only solution, while for $\alpha _0=0$  we get the only restriction on  $m$ and
$n$: $m-n=1$, which means that in the modules for  $(m,n)=(n+1,n)$  for  $n>1$
new primary fields appear when  $\alpha _0 \rightarrow  0$  at the second
level.

In terms of the null-vector condition this means that

\beq
\new
\begin{array}{c}
{\cal L}_{-2}\Phi _{m,n} = [T -
(1+\alpha _0\alpha _+)\partial ^2]e^{i\alpha _{m,n}x} =\\
= \left[ -{1\over 2}(\partial x)^2 + i(\alpha _0+\alpha _{m,n})\partial ^2x -
\right.\\
- \left.
(1+\alpha ^2_0+\alpha _0\sqrt{\alpha ^2_0+2})(-\alpha ^2_{m,n}(\partial x)^2+
i\alpha _{m,n}\partial ^2x)\right] e^{i\alpha _{m,n}x} =\\
= \left[ {1\over 2}\{(m-n)^2-1\}(\partial x)^2\right] e^{i\alpha _{m,n}x} +\\
+
\alpha _0\times \left\lbrace \left\{ \left[ \alpha _0\left( \left(
{m+n+2\over 2}\right) ^2
+\left( {m-n\over 2}\right) ^2\right)  +\right. \right. \right.       \\
+ \left. {1\over 2}\sqrt{\alpha ^2_0+2}
(m+n+2)(m-n)\right] (1+\alpha ^2_0+\alpha _0\sqrt{\alpha ^2_0+2}) + \\
+ \left. {1\over 2}(m-n)^2(\alpha _0+\sqrt{\alpha ^2_0+2})\right\}
(\partial x)^2 +\\
+ (1-\alpha _{m,n}(\alpha _0+\sqrt{\alpha ^2_0+2}))i\partial ^2x\left.
\right\rbrace e^{i\alpha _{m,n}x}\ .
\end{array}
\eeq
For  $\alpha _0 \rightarrow  0$  (and only in this case) this has additional
zeros for  $m-n=1$:

\beq
{\cal L}_{-2}\Phi _{n+1,n} = \alpha _0
\left\lbrace {1\over 2}\sqrt{2}(2n+3)(\partial x)^2\right\rbrace e^{i\alpha _{n
+1,n}x} + O(\alpha ^2_0) \rightarrow  0\ .
\eeq
Therefore

\beq
\|{\cal L}_{-2}\Phi _{n+1,n}\|^2 \sim  \alpha ^2_0
\eeq
for  $N = 2J+1 = 2$, where  $2J = m-n = 1$, and for {\it any} $n$. The only
kind of representation which does not acquire any new null-vectors at the point
$c=1$  has  $n=1$ (in our conventions when $m>n$), and the corresponding
$\tilde \chi _{m,1}(\tau ) = \displaystyle{{1\over
\eta (q)}[q^{\tilde \Delta _{m,1}} -
q^{\tilde \Delta _{m,-1}}]}$  becomes exactly the Kac-Rocha-Caridi characters
(9) with $2J = m-1$.

Our next purpose is to generalize this more or less widely known construction
from the case of $SU(2)$ to other simply laced algebras (only simply
laced algebras are just so simply related to the torus compactifications: in
the case of simply laced algebras the Kac-Moody algebra with $k=1$ possesses an
obvious free-field representation in terms of $r_G$ scalar fields).

\subsection{Torus compactifications}

{\it The origin of the model structure. Relevance of $W_G$-algebra.} Proceed
now to the case of arbitrary simply laced algebra {\it G}. Namely, consider the
set of $r_G = rank(G)$ {\it free} fields  ${\Bf X} = \{X_1,\ldots,X_{r_G}\}$
with Lagrangian
$\partial {\Bf X}\bar \partial {\Bf X}$. Usually the chiral algebra in the
$r_G$-dimensional free field theory is  $\hat U(1)^{r_G}\times \hat U(1)^{r_G}$
and the Virasoro algebra is generated by  $T = \displaystyle{
-{1\over 2}\partial {\Bf X}\partial {\Bf X}}$.
However, now it is reasonable to consider not
only a Virasoro sub-algebra of the chiral-algebra, but larger (higher-spin)
$W_G$-algebra, essentially generated by operators like
$\sum ({\Bf \mu } \partial {\Bf X})^n$. Compactification is defined by
$r$-dimensional
lattice  $\Gamma  = \{{\Bf \gamma } \}$:  ${\Bf X} \sim  {\Bf X} + 2\pi
{\Bf \gamma } $. Naive Virasoro
primaries are  $e^{i{\Bf p}{\Bf x}}e^{i\bar {\Bf p}\bar {\Bf x}}$, where

\beq
\new
\begin{array}{c}
{\Bf p} = {\Bf \gamma } ^\ast  + {1\over 2}{\Bf \gamma }\ ,\\
\bar {\Bf p} = {\Bf \gamma } ^\ast  - {1\over 2}{\Bf \gamma }\ ;\\
{\Bf \gamma } \in \Gamma \hbox{, } {\Bf \gamma } ^\ast \in \Gamma ^\ast\ ,\\
\end{array}
\eeq
where  $\Gamma ^\ast $ is dual lattice to  $\Gamma $, $i.e.$
${\Bf \gamma } {\Bf \gamma } ^\ast  = integer$. However, for particular
compactification on
lattice, the chiral algebra may become  $\hat G_{k=1}$ (of course, any
subalgebra $H \subset  G$  is also allowed), generated by
${\cal J}_{\Bf \alpha} =
e^{i{\Bf \alpha}{\Bf x}}$, ${\cal H}_{\Bf \nu }  = {\Bf \nu } \partial
{\Bf x}$, where  $\alpha $  are all the
roots of $G$ and  ${\Bf \nu } $  form some basis in the Cartan (hyper)plane.
This
happens whenever  $\Gamma$   (${\Bf \gamma } \in \Gamma $) is the
{\it root lattice} of
$G$; then the charges  $\displaystyle{Q_{\Bf \alpha}  = \oint
{\cal J}_{\Bf \alpha} }$
and  $\displaystyle{Q_{\Bf \nu }  =
\oint   {\cal H}_{\Bf \nu } }$, which generate the action of the global
symmetry algebra
$G$ commute with the stress tensor, and the Virasoro primaries should form
representations of {\it G}. This implies, that just as it was in the
$SU(2)$ case along with the ``naive" primaries (or tachyons)
$e^{i{\Bf p}{\Bf x}}e^{i\bar {\Bf p}\bar {\Bf x}}$ there should be also
``non-naive", arising from the
naive ones by the action of {\it G}. However, when $r_G>1$ this is not the
whole story: the set of non-tachyon Virasoro primaries is much bigger. (In
critical string theory this is a well-known phenomenon -- there are much more
primary fields than only tachyons: gravitons and other higher spin
excitations.) Therefore in order to diminish the set of primary fields and make
it related to the {\it model} of $G$ (where every representation appears
exactly once), it is reasonable to consider a set of  $W_G$-primaries.
Operators of  $W_G$-algebra are essentially given by  $\sum _{a=0}
({\Bf \nu } _a\partial {\Bf x})^n$ with  $n = 1,\ldots,r_G$ (${\Bf \nu } _a$
are certain vectors in the
Cartan (hyper-) plane, related to the fundamental weights, see [20]). The
$W_G$-algebra (or its universal envelope) can be defined as a piece of the
chiral algebra (universal envelope of $\hat G_1$ in our case), which commutes
with all the charges $Q_{\Bf \alpha}$, $Q_{\Bf \nu } $. Therefore all
$W_G$-primaries still
need to form multiplets of $G$, and the set of $W_G$-primaries form the
{\it model} $M[G]$. In order to demonstrate that there are no more
$W_G$-primaries (which is not true for example for the only Virasoro-primaries)
we refer to the formulas for the one-loop partition function (just as we did in
the $SU(2)$ case):

$$
{\cal Z}(\tau ,\bar \tau ) = |\eta (q)^{-r_G}|^2
\sum _{{\Bf \nu } \in \Gamma ^\ast /\Gamma } \sum  _{\Bf \epsilon }
\left| \Theta \left[ {{\Bf \nu }  + {\Bf \epsilon} }\atop {{\Bf 0}}\right]
(\tau )\right| ^2
$$
with ${\Bf \epsilon }$ running over set of vectors  $\displaystyle{\{{1\over 2}
{\Bf e}_i\}}$ and
{\Bf 0}
($\{{\Bf e}_i\}$
being the basis of lattice $\Gamma$ ) [14]. The item with
${\Bf \epsilon } = {\Bf 0}$,

\beq
{\cal Z}(\tau ,\bar \tau ) = |\eta (q)^{-r_G}|^2
\sum _{{\Bf \nu } \in \Gamma ^\ast /\Gamma }
\left| \Theta \left[ {{\Bf \nu } }\atop {{\Bf 0}}\right]
(\tau )\right| ^2\hbox{ ,}
\eeq
is modular invariant by itself and is, in fact, a one-loop partition function
of the  $k=1$ WZW model for a simply laced  $G$ [21]. The corresponding
1-loop partition function in ``chiral" or ``open" sector is

\beq
\new
\begin{array}{c}
Z(\tau ) \equiv   \eta (q)^{-r_G} \sum _{{\Bf \nu } \in \Gamma ^\ast /\Gamma }
\Theta \left[ {{\Bf \nu } }\atop {0}\right] (\tau ) =\\
= \sum _{{\Bf \Lambda } \in \Gamma _{\ast}}  D_{\Bf \Lambda } \chi _{\bf
\Lambda } (\tau ).
\end{array}
\eeq
Here we label the highest weight representations  $R_G[{\Bf \Lambda } ]$
of  $G$  by
the highest weight vectors ${\Bf \Lambda } $, lying in a ``positive" Weyl
chamber
$\Gamma ^+$. Dimension of $R_G[{\Bf \Lambda } ]$ is given by a product over all
{\it positive} roots ${\Bf \alpha}$  [22]:

\beq
D_{\Bf \Lambda }  = \prod_{{\Bf \alpha} \in \Delta _+}
{\langle {\Bf \Lambda } +{\Bf \rho } ,{\Bf \alpha} \rangle \over \langle
{\Bf \rho } ,{\Bf \alpha} \rangle }
\eeq
(${\Bf \rho }  = \displaystyle{{1\over 2} \sum _{{\Bf \alpha} \in \Delta_{+}}
{\Bf \alpha} }$,
$\langle $ ,
$\rangle $ is the scalar product in the Cartan (hyper-)plane). According to our
above reasoning the same  ${\Bf \Lambda } $  can be used to label the
$W_G$-primaries, and thus the irreducible representations
${\cal R}_{W_G}[{\Bf \Lambda } ]$ of the  $W_G$-algebra with  $c=r_G$. (Note
that
${\cal R}_{W_G}[{\Bf \Lambda } ]$  is certainly not the same as
$R_G[{\Bf \Lambda } ]$:
these are representations of different algebras!). For the sake of brevity we
use the symbol  $\chi _{\Bf \Lambda } (\tau )$  instead of
$\chi \left[ {\cal R}_{W_G}[{\Bf \Lambda } ]\right] (\tau )$ to denote the
analogues
of the Virasoro Kac-Rocha-Caridi characters of the irreducible representation
${\cal R}_{W_G}[{\Bf \Lambda } ]$ with conformal dimensions
$\Delta _{\Bf \Lambda }  =
\dst{{1\over 2}{\Bf \Lambda } ^2}$:

\beq
\chi _{\Bf \Lambda } (\tau ) =
\eta (q)^{-r_G}\sum _{\sigma \in {\cal W}}\det(\sigma )q^{{1\over 2}
({\Bf \Lambda } +
{\Bf \rho } -\sigma ({\Bf \rho } ))^2}.
\eeq
Here ${\cal W}$ is the finite Weyl group of the algebra $G$ and $\det(\sigma )$
stands for the determinant of transformation  $\sigma \in {\cal W}$, (for
$G=SU(2)$ the Weyl group is  ${\cal W}={\Bf Z}_2$, thus there are only two
items in the sum (24) and we return to eq.(7)). These characters depend only on
dimension  $\Delta $  and therefore are the same for all the primaries
$D_{\Bf \Lambda } $, belonging to representation  $R[{\Bf \Lambda } ]$. Thus
they give equal
contributions to the sum (22), that is why the factors  $D_{\Bf \Lambda } $
appear in
front of corresponding  $\chi$'s. It is eq.(22) that confirms our statement
that  $W_G$-primaries form {\it model} of G. Thus it remains to explain the
origin of eqs.(21-22) and (24).

{\it Proof of eq.(22).} The proof is a bit more subtle but very close to that
of the $SU(2)$ case.

Let us calculate the sums at the $r.h.s.$ of eq.(22). First of all, note, that
these sum can be taken over entire weight lattice. Indeed, as a simple
corollary of eqs.(23) and (24) we have for any  $\sigma \in {\cal W}$  and
weight  ${\Bf \nu } $:

\beq
D_{\Bf \nu } \chi _{\Bf \nu } (\tau ) = D_{{\Bf \nu } _\sigma }
\chi _{{\Bf \nu } _\sigma }(\tau ),
\eeq
where  ${\Bf \nu } _\sigma \equiv  \sigma ({\Bf \nu } ) + \sigma ({\Bf \rho } )
- {\Bf \rho } $. This
implies that for any lattice  ${\cal T}$

\beq
\sum _{{\Bf \nu } \in {\cal T}_{+}} D_{\Bf \nu } \chi _{\Bf \nu } (\tau ) =
{1\over \hbox{ord} {\cal W}}
\left\lbrace  \sum _{{\Bf \nu } \in {\cal T}} D_{\Bf \nu } \chi _{\Bf \nu }
(\tau )\right\rbrace ,
\eeq
where  $\hbox{ord} {\cal W}$  is the order (the number of elements) of Weyl
group,
${\cal T}_+$ is intersection of  ${\cal T}$  with the Weyl chamber, and
$\hat {\cal T}$  is the union of specifically shifted images of  ${\cal T}_+$
under all the transformations from the Weyl group:

\beq
\hat {\cal T} =
\bigcup_ {\sigma \in \cal W}[\sigma ({\cal T}_+) + \sigma ({\Bf \rho } ) -
{\Bf \rho } ].
\eeq
In general,  $\hat {\cal T}$  does not need to coincide with original lattice
${\cal T}$. This is true for {\it root} lattice  $\hat \Gamma = \Gamma $, but
for the case we consider now:  $\hat \Gamma ^\ast  \neq  \Gamma ^\ast $. (In
the simplest example of $SU(2)$  $\Gamma ^\ast = \{n/\sqrt{2}
,n\in {\Bf Z}\}$, $ \rho  = 1/\sqrt{2}$, $\Gamma ^\ast _+ = \{n/\sqrt{2}
,n\in {\Bf Z}$, $n\geq 0\}$ and $\hat \Gamma ^\ast _+ = \{n/\sqrt{2}
,n\in {\Bf Z}$, $n\neq -1\}$. Thus the difference between  $\Gamma ^\ast _+$
and  $\hat \Gamma ^\ast _+$ consists of a single point  $\nu  = -\rho  $, and,
according to (23) this is exactly the point where  $D_{-\rho }= 0$, $i.e.$
which does not contribute to the sum at the $r.h.s.$ of (26), so that it can be
taken over the whole lattice  $\Gamma ^\ast _+$.) The last statement is also
true for all other simply laced algebras  $G$: in general the difference
between  $\Gamma ^\ast _+$ and  $\hat \Gamma ^\ast _+$ is no longer a point,
but consists of hyperplanes of codimension  1, such that for any
${\Bf \nu }  \in
\Gamma ^\ast - \hat \Gamma ^\ast $ the sum  ${\Bf \nu } +{\Bf \rho } $  is
orthogonal at
least to one of the positive roots and thus, according to (23) the
corresponding  $D_{\Bf \nu }  = 0$, and the sum at the $r.h.s.$ of (26)
can be taken
over entire  $\Gamma ^\ast $ instead of  $\hat \Gamma ^\ast $. We conclude that
(up to a factor $\hbox{ord} {\cal W})$ the sum at the $r.h.s.$ of eq.(26) is
over
entire lattices  $\Gamma ^\ast $.

Now the calculation can be done as follows. First, we have

\beq
\new
\begin{array}{c}
Z(\tau ) = \sum _{{\Bf \nu } \in \Gamma ^\ast _+}D_{\Bf \nu } \chi _{\Bf \nu }
(\tau ) =
{1\over \hbox{ord} {\cal W}}
\sum _{{\Bf \nu } \in \hat \Gamma ^\ast }D_{\Bf \nu }
\chi _{\Bf \nu } (\tau )
=\\
= {1\over \hbox{ord} {\cal W}} \sum _{{\Bf \nu } \in \Gamma ^\ast }D_{\Bf \nu }
\chi _{\Bf \nu } (\tau )\ ,
\end{array}
\eeq
when, we have also used that  $D_{\Bf \nu } =0$  for  ${\Bf \nu }  \in
\Gamma ^\ast -\hat \Gamma ^\ast $. Substituting (28), and changing the
summation variable  ${\Bf \Lambda }  = {\Bf \nu }  + {\Bf \rho }  -
s({\Bf \rho } )$  one gets

\beq
\new
\begin{array}{c}
\eta (q)^{r_G} Z(\tau ) ={1\over \hbox{ord} {\cal W}}
\sum _{{\Bf \nu } \in \Gamma ^\ast }
\sum _{s\in {\cal W}} \prod_{{\Bf \alpha} \in \Delta _+}
{\langle {\Bf \nu } +{\Bf \rho } ,{\Bf \alpha} \rangle \over \langle
{\Bf \rho } ,{\Bf \alpha} \rangle }
\det (s) q^{{1\over 2}[{\Bf \nu } +{\Bf \rho } -s({\Bf \rho } )]^2} =\\
= {1\over \hbox{ord} {\cal W}} \sum _{s\in {\cal W}} \det (s)
\sum _{{\Bf \Lambda } \in \Gamma ^\ast} \prod_{{\Bf \alpha} \in \Delta _+}
{\langle {\Bf \Lambda } +s({\Bf \rho } ),{\Bf \alpha} \rangle \over \langle
{\Bf \rho } ,{\Bf \alpha} \rangle }
q^{{1\over 2}{\Bf \Lambda } ^2} =\\
= \sum _{{\Bf \Lambda } \in \Gamma ^\ast } q^{{1\over 2}{\Bf \Lambda } ^2}
\end{array}
\eeq
(we do not need to take care about boundary terms since we sum now over
{\it entire} lattice  $\Gamma ^\ast )$, and it has been used that  $s^2 = 1$
for  $s\in {\cal W}$, as well as:

\beq
\sum _{\sigma \in {\cal W}} \det(\sigma )
{\langle {\Bf \Lambda } +\sigma ({\Bf \rho } ),{\Bf \alpha}
\rangle \over \langle {\Bf \rho } ,{\Bf \alpha}
\rangle } = \hbox{ord} {\cal W}
\eeq
for any  ${\Bf \Lambda } $. Finally, the $r.h.s.$ of eq.(29) is equal to

\beq
\sum _{{\Bf \nu } \in \Gamma ^{\ast}}  q^{{\Bf \nu } ^2/2} =
\sum _{{\Bf \nu } \in \Gamma ^\ast /\Gamma }
\left\lbrace  \sum _{{\Bf \lambda } \in \Gamma } q^{({\Bf \lambda } +
{\Bf \nu } )^2/2}\right\rbrace
= \sum _{{\Bf \nu } \in \Gamma ^\ast /\Gamma } \Theta \left[
{{\Bf \nu } }\atop {{\Bf 0}}\right] (\tau )
\eeq
where the {\it lattice}  $\Theta $-function is defined as a sum over the
{\it root} lattice  $\Gamma $. This proves eq.(22) provided (24) is correct.

{\it Characters (24) as the limit of Fateev}-{\it Lukyanov characters.}
In order to derive the formula (24) for the characters of irreducible
$W_G$-representations we will just apply the same trick as we demonstrated in
the case of  $SU(2)$. Namely, let us make use of the known
$W_G$-characters for the ``minimal" series, arising for certain values of
$c <r_G$

\beq
c = r_G - 12\alpha ^2_0{\Bf \rho } ^2 = r_G - 6{(p-p')^2\over pp'}{\Bf \rho }
^2
\eeq
and then take the limit $c \rightarrow  r_G$, or  $\alpha _0 \rightarrow  0$.
According to [23]

\beq
\new
\begin{array}{c}
\chi _{{\Bf \Lambda } _1,{\Bf \Lambda } _2}(\tau ) = \eta (q)^{-r_G}
\sum _{s_1,s_2\in {\cal W}} {\det (s_1)\det (s_2)\over \hbox{ord} {\cal W}}
\Theta \left[ {p s_1 {\Bf \Lambda } 1 -p's_2
{\Bf \Lambda } 2}\atop {{\Bf 0}}\right] (pp'\tau ) =\\
= \eta (q)^{-r_G} \sum _{s\in {\cal W}}\det (s) \sum _{{\Bf \alpha}
\in \Gamma }
\exp \left( {i\pi \tau \over pp'}(ps{\Bf \Lambda } _1 - p'{\Bf \Lambda } _2 -
2pp'{\Bf \alpha} )^2\right)
\end{array}
\eeq
After one takes the limit  $p \rightarrow  \infty $, $p' \rightarrow  \infty $
so that  $p'-p$  is fixed and finite only the term with
${\Bf \alpha} ={\Bf 0}$  from the
whole sum over root lattice survives in (33). After a conventional redefinition
${\Bf \Lambda } _i \rightarrow  {\Bf \Lambda } _i + {\Bf \rho } $  and setting
${\Bf \Lambda } _1 = {\Bf 0}$
(the analog of  $n=1$  condition in the  $SU(2)$  case) and
${\Bf \Lambda } _2 \equiv  {\Bf \Lambda } $ one recognizes the formula (24).

\subsection{Virasoro and  $W$-primaries from BRST formalism}

There exists a somewhat different way to observe the complicated structure of
primary fields in  $c = 1$  and  $c = r_G$ string theories (see for example
[2], another example was in fact given in sect.2.1. above  when we discussed
the occurrence of new null-vectors in the limit of  $c = 1$). Since it
illustrates some new features of these theories we mention this kind of ideas
here without going into many details. Let us start from uncompactified theory
where any (and not essentially rational) values of momenta are admissible. For
generic momenta (and dimension  ${\Bf p}^2/2$) there is a single Virasoro or
$W$-primary:  $\Phi _{\Bf p} = e^{i{\Bf p}{\Bf X}}$. Occurrence of non-trivial
primaries for
certain values of  ${\Bf p}$  can be observed as follows. Apply some algebraic
combination of Virasoro and/or $W$-operators to  $\Phi _{\Bf p}$ and look
at the
norm of the resulting operator,
$\|{\cal P}^{({\Bf p})}_{-N}\Phi _{\Bf p}\|^2$ as a
function of  ${\Bf p}$. If this function is vanishing at the point
${\Bf p}_o$, one can
introduce a new field of non-vanishing norm,  $
\lim_{{\Bf p} \rightarrow {{\Bf p}_o}} \{|{\Bf p}-{\Bf p}_o|^{-1}{\cal P}
^{({\Bf p})}_{-N}\Phi _{\Bf p}\}$, which is a new non-naive
primary field. We present an example of this phenomenon (to be added to those
already given in [2])\footnote{Note that
this mechanism of generating new primaries as null-vectors is not
the only possible. In fact we consider cokernel of the chiral algebra. If we
discuss Virasoro (rather than appropriate $W$-) algebra for  $r_G>1$  the
cokernel is very large and there is plenty of Virasoro primaries which are
neither tachyonic operators, nor can be described as null-vectors, as we have
in critical string theory.}.

First, we demonstrate how it works for  $c = 1-12\alpha ^2_0$ and illustrate
the occurrence of non-naive Virasoro primaries at appropriate values of
momentum  $p$. Again we shall discuss only the simplest non-trivial example of
the null-vector at level  $N=2$. Consider the Virasoro null-vector in the
module of  $e^{ipX}$. The non-positive Virasoro generators act as

\beq
\new
\begin{array}{c}
L_0e^{ipX} \sim  (p^2/2 - \alpha _0p)e^{ipX};\\
L_{-1}e^{ipX} \sim  ip\partial X\ e^{ipX};     \\
L_{-2}e^{ipX} \sim  \left[ i(p+\alpha _0)\partial ^2X -
{1\over 2}(\partial X)^2\right] e^{ipX}              ;   \\
L^2_{-1}e^{ipX} \sim  [ip\partial ^2X - p^2(\partial X)^2]e^{ipX}.
\end{array}
\eeq
Then

\beq
{\cal L}^{(p)}_{-2}e^{ipX} = \{L_{-2}+aL^2_{-1}\}e^{ipX} =
\left\{ i[p-\alpha _0+ap]\partial ^2X + \left[ -{1\over 2} - ap^2\right]
(\partial X)^2\right\} e^{ipX}.
\eeq
To get in (35) zero, one has to put

\beq
\new
\begin{array}{c}
ap = -(p-\alpha _0)\ ,\\
p^2 = - {1\over 2a}\ .
\end{array}
\eeq
The formulae (36) certainle coincide with usual second level null-vector
conditions  (see, for example [24])

\beq
\new
\begin{array}{c}
c = 1-12\alpha ^2_0 = 2\Delta (5-8\Delta )/(2\Delta +1) = 1 -
(4\Delta -1)^2/(2\Delta +1);\\
a = - 3/2(2\Delta +1)\hbox{; } \alpha _0 = (4\Delta -1)/2\sqrt{3(2\Delta +1)}
\end{array}
\eeq
for  $\Delta  = p^2/2 - \alpha _0p$.

Now we shall turn to the example which is the simplest possible, involving
higher $W$-operators. Let us take  $G = SU(3)$, $c = 2$ ($\alpha _0 = 0$),
level
$N=2$. Then

\beq
T = - {1\over 2} \sum _{\Bf \nu }  ({\Bf \nu } \partial {\Bf X})^2 = -
{1\over 2}(\partial {\Bf X})^2\hbox{; }     W = \sum _{\Bf \nu }  ({\Bf \nu }
\partial {\Bf X})^3.
\eeq
The Virasoro and  $W$-modes act as

\beq
\new
\begin{array}{c}
L_{-1}e^{i{\Bf p}{\Bf X}} = i{\Bf p}\partial {\Bf X}e^{i{\Bf p}{\Bf X}};\\
L_{-2}e^{i{\Bf p}{\Bf X}} = \left[ i{\Bf p}\partial ^2{\Bf X} -
{1\over 2}(\partial
{\Bf X})^2\right] e^{i{\Bf p}{\Bf X}};\\
L^2_{-1}e^{i{\Bf p}{\Bf X}} = \left[ i{\Bf p}\partial ^2{\Bf X} -
({\Bf p}\partial
{\Bf X})^2\right] e^{i{\Bf p}{\Bf X}}.
\end{array}
\eeq
Now the only parameter  {\it a}  is not enough ensure vanishing of the sum
$\{L_{-2}+aL^2_{-1}\}e^{i{\Bf p}{\Bf X}}$ for any  ${\Bf p}$. Thus in order
to get null-vectors
we need to consider ``mixed" Virasoro--$W$  action ($i.e.$ there is no
null-vector built from only the Virasoro modes). For the  $W$-operators one
has:

\beq
\new
\begin{array}{c}
W_{-1}e^{i{\Bf p}{\Bf X}} = 3 \sum _{\Bf \nu }  ({\Bf \nu } \partial {\Bf X})
(i{\Bf p}{\Bf \nu } )^2e^{i{\Bf p}{\Bf X}};\\
W_{-2}e^{i{\Bf p}{\Bf X}} \sim  3 \sum _{\Bf \nu }  [({\Bf \nu } \partial ^2
{\Bf X})(i{\Bf p}{\Bf \nu } )^2 +
({\Bf \nu } \partial {\Bf X})^2(i{\Bf p}{\Bf \nu } )]e^{i{\Bf p}{\Bf X}};\\
W^2_{-1}e^{i{\Bf p}{\Bf X}} \sim  9
\sum _{{\Bf \nu } ,{\Bf \nu } '}({\Bf \nu } \partial {\Bf X})^2({\Bf \nu }
{\Bf \nu } ')(i{\Bf p}{\Bf \nu } ')^2e^{i{\Bf p}{\Bf X}};\\
L_{-1}W_{-1}e^{i{\Bf p}{\Bf X}} \sim  3 \sum _{\Bf \nu }  [({\Bf \nu }
\partial ^2{\Bf X})(i{\Bf p}{\Bf \nu } )^2 +
({\Bf \nu } \partial {\Bf X})(i{\Bf p}\partial {\Bf X})(i{\Bf p}{\Bf \nu } )^2]
e^{i{\Bf p}{\Bf X}}.
\end{array}
\eeq
The formulae for the particular case of  $SU(3)$  read:

\beq
\new
\begin{array}{c}
W \sim  (\partial X_1)^3 - 3(\partial X_1)(\partial X_2)^2;\\
{1\over 3}W_{-1}e^{i{\Bf p}{\Bf X}} \sim  [(p^2_2-p^2_1)\partial X_1 +
2p_1p_2\partial X_2]e^{i{\Bf p}{\Bf X}};\\
{1\over 3}L_{-1}W_{-1}e^{i{\Bf p}{\Bf X}} \sim  [(p^2_2-p^2_1)\partial ^2X_1 +
2p_1p_2\partial ^2X_2 + \\
+ \{(p^2_2-p^2_1)\partial X_1 +
2p_1p_2\partial X_2\}(ip\partial X)]e^{i{\Bf p}{\Bf X}};\\
{1\over 9}W^2_{-1}e^{i{\Bf p}{\Bf X}} \sim  \{[(p^2_2-p^2_1)\partial X_1 +
2p_1p_2\partial X_2]^2 +\\
+ [(p^2_2-p^2_1)[(\partial X_1)^2-(\partial X_2)^2] -
4p_1p_2(\partial X_1\partial X_2)] +\\
+ [2(p^2_2-p^2_1)[ip_1\partial ^2X_1 - ip_2\partial ^2X_2] -
4p_1p_2[ip_2\partial ^2X_1 + ip_1\partial ^2X_2]]\}e^{i{\Bf p}{\Bf X}};\\
{1\over 3}W_{-1}e^{i{\Bf p}{\Bf X}} \sim  \{[(p^2_2-p^2_1)\partial ^2X_1 +
2p_1p_2\partial ^2X_2] + ip_1[(\partial X_1)^2-(\partial X_2)^2]\hbox{ -}\\
- 2ip_2(\partial X_1\partial X_2)\}e^{i{\Bf p}{\Bf X}}.
\end{array}
\eeq
Combining (41), we see that

\beq
[L_{-2}-L^2_{-1}]e^{i{\Bf p}{\Bf X}} = \left[ ({\Bf p}\partial {\Bf X})^2-
{1\over 2}
(\partial {\Bf X})^2\right] e^{i{\Bf p}{\Bf X}}.
\eeq
and

\beq
\new
\begin{array}{c}
[(\alpha /3)L_{-1}W_{-1} + (\beta /9)W^2_{-1} + (\gamma /3)W_{-2}]
e^{i{\Bf p}{\Bf X}} \sim \\
\sim  \left\lbrace \partial ^2X_1[\alpha (p^2_2-p^2_1) + \gamma (p^2_2-p^2_1) +
6\beta (p^2_2-p^2_1)p_1] +\right.\\
+ \partial ^2X_2[2\alpha p_1p_2 + 2\gamma p_1p_2 - 6\beta (p^2_2-p^2_1)p_2] +\\
+ (\partial X_1)^2[i\alpha p_1(p^2_2-p^2_1) - i\gamma p_1] +
(\partial X_1)^2[2i\alpha p_1p^2_2 + i\gamma p_1] +\\
+ \partial X_1\partial X_2[i\alpha p_2(p^2_2-p^2_1) + 2i\alpha p^2_1p_2 +
2i\gamma p_2]\left. \right\rbrace e^{i{\Bf p}{\Bf X}}\ .
\end{array}
\eeq
The solution to the null-vector conditions (43) exists, when

\beq
\beta =0\hbox{, }   \gamma  = -\alpha   i\alpha  = \sqrt{6}/4.
\eeq
(in our conventions  $X_1$ is chosen along $\bar {\Bf \mu }  = (0,\sqrt{2/3}))$
and
looks like

\beq
\new
\begin{array}{c}
p_1 = \pm 1/\sqrt{6}\hbox{; } p_2 = \pm 1/\sqrt{2};\\
p_1 = \pm \sqrt{2/3}\hbox{; } p_2 = 0
\end{array}
\eeq
and corresponds to six fundamental weights of  {\bf 3}  and  $\bar {\bf 3}$.

\section{Coupling to $W_G$-gravity}

We demonstrated in the previous section that the  $W_G$-primaries in the matter
theory, compactified on the root lattice of the Lie algebra  $G$, form the
{\it model} of G. However, the spectrum of this matter theory which is just an
ordinary conformal theory is by no means exhausted by the primaries: there are
many $W_G$-descendants, and these do not respect in any obvious way the
structure of the {\it model}, which is of the main interest for us. Therefore,
if we want to distinguish this {\it model} structure we need to get rid of
descendants. The simplest way to do this is just to gauge the $W_G$-algebra,
$i.e.$ consider the  $W_G$-string model: by coupling to  $W_G$-gravity we will
cancel the contribution of all the  $W_G$-descendants to the physical
correlation functions\footnote{It
means that Liouville and ghost contributions just cancell Dedekind
$\eta $-factor in (22).}.
In this short section we demonstrate that this transition, which essentially
involves dressing of matter primary fields by appropriate  $W$-Toda fields,
indeed preserves the  $G$-{\it model} structure, which therefore becomes the
feature of the set of observables in this string theory (again we will consider
a kind of ``chiral" theory). The reason for this is very simple: all the
$W$-Toda fields just commute with the action of  $G$-charges  $Q_{\Bf \alpha}
$,
$Q_{\Bf \nu } $, and thus dressing does not affect any structures, related to
the
algebra  $G$ (but may give a correct selection rule).

The subject of  $W$-strings (or closely related  $W$-gravity) has been already
discussed in the literature [8-13]. Attempts were made in order to work out an
analogue of the DDK approach (or [25]). It has been also
confirmed by the BRST calculation that observable operators, obtained in this
way, are indeed annihilated by BRST charge [26]. Of course the DDK-like
construction, which assumes an oversimplified dependence on the ghost and
Liouville fields, does not need to give {\it all} the observables\footnote{Just
as it happens in the $SU(2)$ case, where along with ``dimension
zero, ghost number one" (or ``dimension one, ghost number zero") operators
implied by the DDK formalism and considered in [1], there are also less trivial
``dimension zero, ghost numbers zero and two" observable operators [7,1,2].
Another example is provided by minimal $c<1$ series, where (at least in some
versions of the formalism) Liouville field does not usually appear in the pure
exponential form (as Virasoro primary in the Liouville sector) as it is implied
in what we call the DDK approach.}.
But even in the sector which it does describe, the DDK approach is less
conclusive than in the case of ordinary strings. The main difference arising in
consideration of  $W_G$-gravity as compared to the ordinary
``Virasoro gravity", is that observable operators are no longer represented as
integrals of (at least any simple) ghost free operators of dimension one.
Instead the adequate ghost free operators have dimension  $\Delta ^{\{G\}} =
2{\Bf \rho } ^2 = \dst{{1\over 6} C_V[G] \dim G}$.
Even triple correlators of such observables
involve non-trivial correlators of ghosts (the moduli space of  $W_G$-gravity
is already non-trivial for a sphere with 3 punctures), which need a reasonable
understanding of  $W$-geometry in order to be defined. This is the main
obstacle on the way to work out the structure of algebra of observables
(reminiscent of operator product expansion) and thus to understand the points
3)--6) of the KPW construction, as described in the Introduction. While
resolution of these problems is beyond the scope of this paper, in this section
we present a bit more details about DDK formalism for $W$-strings to illustrate
what we just said about it and also to demonstrate that it preserves the
{\it model} structure of our  $c = r_G$ matter theory from the previous
section.

The analogue of conventional Liouville action is provided by conformal
$G$-Toda action:

\beq
\int _{d^2z}
\left\{ |\partial {\Bf \phi} |^2 + \beta _0{\cal R}{\Bf \rho } {\Bf \phi}  +
\sum _{i=1}
\eta _ie^{{\Bf \alpha} _i{\Bf \phi} }\right\} ,
\eeq
the sum is over  $r_G$ {\it simple} roots of  {\it G}. In the framework of the
KPW construction we consider a point where all  $\eta _i=0$. Besides the
$r_G$-component  $W$-Toda field  ${\Bf \phi} $  there are also  $r_G$ species
of
ghosts: Grassmann $b$,$c$-systems with spins  $j\in S_G$ with the first-order
action

\beq
\int _{d^2z} \sum _{j\in S_G} \{b_j\bar \partial c_{1-j} + c.c.\}.
\eeq
Occurrence of ghosts is obvious from the geometrical meaning of  $W$-fields
(describing certain flag structures in the jet bundle over Riemann surface
[10]) and, more technically, from interpretation of (12) as a result of the
Drinfeld-Sokolov reduction of WZW model [27-30]. The set  $S_G$ is nothing but
the set of  $G$-invariants or the Casimir orders, which is different for the
three series $A$, $D$ and $E$  of the simply laced algebras: for  $SU(r+1)$ --
$j=2,\ldots,r_G+1$  (and $1-j = -1,\ldots,-r_G$);  $SO(2r)$ --
$j = 2,4,\ldots,2r-2$  and
$r$;  $E_6$ --  $j = 2,5,6,8,9,12$;
$E_7$ -- $j=2,6,8,10,12,14,18$; and for  $E_8$ -- $j=
2,8,12,14,18,20,24,30$. Thus the central charge of ghosts is equal to

\beq
c_{ghosts} =  \sum _{j\in S_G} [-2(6j^2-6j+1)] = - 48{\Bf \rho } ^2 -
2r_G\hbox{.}
\eeq
The central charge of the  $W$-Toda fields equals  $c_{\Bf \phi}  = r_G +
48\beta ^2_0{\Bf \rho } ^2$. Since the total central charge  $c_{matter} +
c_{\Bf \phi}  +
c_{ghosts} = 0$, we have:  $48(\beta ^2_0-1){\Bf \rho } ^2 = c_{matter} - r_G$,
and,
therefore, for  $c_{matter} = r_G -  \beta _0= \pm 1$.

The naive DDK approach implies the following algorithm for building up the
observables in the  $W_G$-string model, which arises by gauging  $W_G$-symmetry
of the matter sector.

{\bf A)} Pick up any  $W_G$-primary from the matter sector. In our particular
model
from sect.2 it is labeled by two indices:  $\Psi _{{\Bf \nu } ,{\Bf \xi } }
({\Bf x}) =
\prod^{r_G}_{i=1} (Q_{-{\Bf \alpha} _i})^{({\Bf \mu } _i{\Bf \xi } )}
\Psi _{{\Bf \nu } ,{\Bf 0}}$,  $\Psi _{{\Bf \nu } ,{\Bf 0}} =
e^{i{\Bf \nu } {\Bf x}}$, where  ${\Bf \nu }$  ($a$ is a
vector in the weight lattice) labels
representations  $R_{\Bf \nu } [G]$  of  $G$
(${\Bf \nu } $ is just its highest weight), while
${\Bf \xi }  \equiv  {\Bf \xi } _R$ labels the element of this representation.
The conformal
dimension  $\Delta _{{\Bf \nu } ,{\Bf \xi } } = {\Bf \nu } ^2/2$  is
independent of  ${\Bf \xi } $.

{\bf B)} Dress this matter field by the  $W$-Toda exponent:
$\Xi _{{\Bf \nu } ,{\Bf \xi } }({\Bf x},{\Bf \phi} ) = \Psi _{{\Bf \nu } ,
{\Bf \xi } }({\Bf x})e^{{Bf \beta} _{\Bf \nu } {\Bf \phi} }$, so that
$\Xi _{{\Bf \nu } ,{\Bf \xi } }$ has the fixed dimension  $\Delta ^{\{G\}}$.
This requirement
restricts the value of ${\Bf \beta }_{\Bf \nu } $:

\beq
\Delta _{{\Bf \nu } ,{\Bf \xi } } - {1\over 2}{\Bf \beta} ^2_{\Bf \nu }  -
2{\Bf \beta }_0
\beta _{\Bf \nu } {\Bf \rho }   =
\Delta ^{\{G\}},
\eeq
or, in our particular model with  $\Delta _{{\Bf \nu } ,{\Bf \xi } } =
\dst{{1\over 2}{\Bf \nu } ^2}$
and
$\beta _0= 1$,

\beq
{1\over 2}{\Bf \nu } ^2 = {1\over 2}({\Bf \beta }_{\Bf \nu }  +
2{\Bf \rho } )^2 + (\Delta ^{\{G\}}-
2{\Bf \rho } ^2).
\eeq
Of course, there are many solutions of this 1-component (scalar) equation for
the $r_G$-component (vector)  ${\Bf \beta }_{\Bf \nu } $ once  ${\Bf \nu } $
is
fixed. However, it
is clear that a distinguished set of solutions arises, if

\beq
\Delta ^{\{G\}} = 2{\Bf \rho } ^2
\eeq
and

\beq
{\Bf \beta }_{\Bf \nu }  = {\Bf \nu }  - 2{\Bf \rho } .
\eeq
(about possible choices of sign see below)\footnote{Such
distinguished choice of  $W$-Toda dressing, corresponding to a
particular solution of the equation, as well as the ghost dressing below, has
been already proposed, for example, in [12].}.

{\bf C)} Add extra ghost factor in order to compensate for non-vanishing
$\Delta ^{\{G\}}$ and create an operator of dimension zero.

In conventional Liouville theory ($2d$ gravity)  $\Delta ^{\{SU(2)\}}= 1$ and
it
is enough to multiply  $\Xi (x,\phi )$  by an ordinary reparameterization ghost
$c_{-1} \equiv  c$  to create an observable operator

\beq
{\cal O}_{\nu  , \xi  }(x,\phi ,c) = \Xi _{\nu  , \xi }(x,\phi )c_{-1} =
\psi _{ \nu ,\xi  }(x)e^{\beta _\nu \phi }c_{-1}.
\eeq
Correlators of observables are evaluated with additional insertions of the form

\beq
\prod^{N^{(2)}}_{\alpha=1} \int_{d^2z} b_2\mu ^{(2)}_\alpha,
\eeq
where  $\mu ^{(2)}_\alpha $ are the Beltrami differentials, associated with the
moduli of complex structure of the surface, and  $N^{(2)}$ is dimension of the
module space. An alternative and conceptually simpler description of
essentially the same operator as (53) in Liouville theory may be given without
any reference to ghosts and  BRST formalism: observable may be defined as

\beq
\hat {\cal O}_{\nu , \xi  }(x,\phi ) = \int _{dz} \Xi _{ \nu  ,\xi  }(x,\phi )
=
\int _{dz} \psi _{ \nu ,\xi  }(x)e^{\beta _\nu  \phi },
\eeq
$i.e.$ as an integral of ghost free operator of dimension one. When the
correlators are evaluated in this representation the integration contours are
deformed and integrals automatically pick up all contributions from the points
where other operator are inserted and as well as from the handles on the
surface. Equivalence between correlators, evaluated in the two representations
(53) and (55) may be achieved by explicit calculation of the ghost contribution
when using representation (53) -- then the integrals over moduli corresponding
to the marked points on Riemann surface turn into the contour- (surface-)
integral of representation (55).

However, again for $G \neq  SU(2)$ the situation is much more complicated. No
natural representation of observables of the form (63) is available (at least
at the moment), so one needs to rely upon a less transparent BRST formulation,
analogous to (53). Generalization is obvious: instead of dressing
$\Xi ({\Bf x},{\Bf \phi} )$  by multiplication with a single ghost field
$c_{-1}$, now one
needs to multiply with a whole combination of ghosts\footnote{We
use the particular dressing by ghost fields -- the contribution of
any  $c_{1-j}$ to this dressing is given by the lowest dimensional operators.
This is consistent with the BRST analysis and properly describe the spectrum of
the theory but does not mean that the operators of other ghost numbers
(pictures) do not make contributions to the correlators.}:

\beq
\new
\begin{array}{c}
{\cal O}_{{\Bf \nu } ,{\Bf \xi } }({\Bf x},{\Bf \phi} ,c) = \Xi _{{\Bf \nu } ,
{\Bf \xi } }({\Bf x},{\Bf \phi} )
\prod_{j \in S_j}
\left\{ c_{1-j}\partial c_{1-j}\partial ^2c_{1-j}\ldots
\partial ^{j-2}c_{1-j}\right\} =\\
= \psi _{{\Bf \nu } ,{\Bf \xi } }({\Bf x})e^{({\Bf \nu } -2{\Bf \rho } ){\bf
\phi} }e^{i\sum (j-1)\varphi _j}.
\end{array}
\eeq
In the last equation we substituted our explicit formula (52) for
${\Bf \beta }_{\Bf \nu } $ and ``bosonized" ghost fields, $i.e.$  $b_j =
e^{-i\varphi _j}$,
$c_{1-j} = e^{i\varphi _j}$. The bosonized ghost action (47) is

\beq
\int _{d^2z} \sum _{j\in S_G} \{b_j\bar \partial c_{1-j} + c.c.\}
= \int _{d^2z} \sum _{j\in S_G}
\left[ \partial \varphi _j\bar \partial \varphi _j +
\imath(j - {1\over 2})\varphi _j{\cal R}\right] .
\eeq
The peculiar combination
$\{c_{1-j}\partial c_{1-j}\partial ^2c_{1-j}\ldots \partial ^{j-2}c_{1-j}\} =
:(c_{1-j})^{j-1}: = e^{i(j-1)\varphi _j}$ has dimension  $\Delta _j=
-j(j-1)/2$, and the entire product in (64) is of dimension

\beq
\sum _{j\in S_G} \Delta _j = {1\over 24} \sum _{j\in S_G} [-2(6j^2-6j+1) + 2] =
{1\over 24}(c_{ghosts} + 2r_G) = -2{\Bf \rho } ^2.
\eeq
Therefore the observable operator indeed has vanishing dimension, but instead
it is of incredibly large ghost charge. This ghost charge is compensated, when
correlators of observables are evaluated, by a product

\beq
\prod _{j \in S_G} \left\lbrace
\prod^{N^{(j)}}_{\alpha=1} \int _{d^2z} b_j\mu ^{(j)}_\alpha \right\rbrace ,
\eeq
which now involves Beltrami differentials associated with all the ``moduli of
$W$-structure". Unfortunately, the meaning of this last notion is much worse
understood than its  $W_2$-counterpart, which is known to be associated with
moduli of the complex structure of the surface (so, that the Beltrami
differentials  $\mu ^{(2)}$, associated with punctures, just describe the shift
of the puncture position). As explained in [10] a holomorphic (chiral) fragment
of $W$-structure is associated with a certain flag in the jet bundle over the
surface. However, there are still unresolved serious difficulties in
understanding the entire (non-chiral) structure\footnote{There
are at least two distinct(?) problems. One is related to the Serre
relations: the Borel subalgebra is not {\it free} for  $G\neq SU(2)$, this
leads to additional non-linear constraints and, in particular to problems with
quantization (see [10,29]). Another problem is that while the stress tensors in
conformal field theory is essentially holomorphic: $T_{\mu \nu } =
\{T_{zz},
T_{z \bar z}, T_{\bar z \bar z}\}$ and $T_{z \bar z}
= 0$, there are no a priori reasons why , say, $W_{zz \bar z}$-component of
the spin-3  $W_3$-operator should be neglected, though no direct
information about it is contained in the structure of chiral $W$-algebra, which
involves only $W_{zzz}$. Moreover, the existing geometric interpretation of
$W_G$-gravity [10]
involves only chiral (holomorphic or antiholomorphic) structures on the
surface.}.
The non-trivial  $``W$-moduli" appear already on the 3-punctured sphere, thus
turning even evaluation of 3-point functions and thus the algebra of
observables into a problem. Resolution of this problem is crucial for
proceeding with the KPW construction in our $c = r_G$ model. We hope to return
to these intriguing problem elsewhere.

The assertion that the operator (56) is indeed a proper observable operator in
$W_G$-string  $c=r_G$ model is somehow confirmed by the fact, that it belongs
to the BRST cohomology. It was explicitly checked in [26] for the case of
$G=SU(3)$. The general form of the BRST operator

\beq
\new
\begin{array}{c}
Q_{BRST} = \sum \ T_a c^a - {1\over 2}\sum \ f^a_{bc}b_ac^bc^c\ ,\\
\left[ T_a,T_b\right] = f^c_{ab}T_c
\end{array}
\eeq
survives even in the case of non-Lie  $W$-algebra with quadratic commutation
relations, if one substitutes to (60) the dependent on generators  $T_a$
structure constants  $f^a_{bc} = f^a_{bc}(T_a)$. For the case of
$SU(3)$  the structure constants are known from [31] and the BRST
operator

\beq
\new
\begin{array}{c}
Q_{BRST} =  -\Delta c^{(-1)}_0 + \sum \ L_mc^{(-1)}_{-m} - {1\over 2} \sum
(m-n):b^{(2)}_{m+n} c^{(-1)}_{-m}c^{(-1)}_{-n}: -\\
- \sum  (2m-n):b^{(3)}_{m+n}c^{(-1)}_{-m}c^{(-2)}_{-n}: -
\sum \ W_mc^{(-2)}_{-m} - {1\over 2}\beta \ \sum
(m-n)L_{-p}b^{(2)}_{m+n+p}c^{(-2)}_{-m}c^{(-2)}_{-n} - \\
- {1\over 2}\gamma \sum  (m-n)\left( {1\over 15}(m+n+2)(m+n+3) -
{1\over 6}(m+2)(n+2)b^{(2)}_{m+n}c^{(-2)}_{-m}c^{(-2)}_{-n}\right)\\
\end{array}
\eeq
(with  $\beta  = \dst{{16\over 22+5c}}$,  $\gamma  = \dst{{5c\over 1044}}$)
satisfies
nilpotency condition for  $c = 100$,  $\Delta  = 2{\Bf \rho } ^2 = 4$ [26].
The form
of the similar BRST operator for any simply laced  $G$  can be got from (60) by
substitution of the structure constants (depending on the generators) of the
quantum algebra  $W_G$. Note that these arguments confirm also our suggested
choice of $W$-Toda dressing of eq.(52).

\section{Conclusion}

To conclude, we discussed in some details the  $G$-induced structure on (a
subsector of) the space of observables in the  $c = r_G$ $W$-string model,
arising after gauging the  $W_G$-symmetry of the 2-dimensional conformal matter
theory, which describes compactification on Cartan torus of a simply laced Lie
algebra  {\it G}. The chiral component of this space may be considered as a
{\it model} of  $G$.

This implies that the result of the KPW construction in this case should be
some algebraic structure, intimately related to the {\it model}, which still
remains to be discovered. This algebraic structure should be related to the
algebra of observables of the theory, $i.e.$ algebra of OPE of physical
operators modulo descendants, or algebra of correlation functions. A systematic
approach to this problem requires a better understanding of the
$W_G$-geometry, what seems to be also an intriguing subject. In particular it
would be nice to have a simple generalization of DDK prescription, allowing one
to represent the physical operators involved into correlation functions in a
simple form. However, a lot of questions on this way need to be understood much
more clear. Finally, we will try to present here at least some of them:

{\bf a)} The relation between algebra of commutators of chiral operators
(contour
integral) and correlation functions (surface integrals) should be more clear
even in the $SU(2)$  case ({\it a priori} they have different
symmetry properties even under the permutation of operators).

{\bf b)} The important remaining problem is complete classification
in spirit of
[7] the physical spectrum of $W_G$-gravity. In fact, be this classification
known a lot of questions we addressed to in the paper will be surely
immediately tractable.

{\bf c)} Even considering of three-point correlation functions may not be
enough
for higher groups  $G$  and higher  $W_G$-gravities, though even the
commutation relation for  $W_G$-algebra itself are quadratic, one could expect
the necessity to consider four-point correlation functions in order to
determine a non-Lie algebraic structure.

{\bf d)} After Witten's proposal [2] it is natural to ask whether any algebra
of
observables can be decomposed into the commutative (ground) ring and
antisymmetric chiral algebra, acting on this commutative ring by derivatives,
and {\it what} is the action of this antisymmetric chiral algebra (see also the
item {\bf f})).

{\bf e)} The next question rather concerns the  $c < 1$  string models. It
would be
nice to find various reductions of the  $W_\infty $-algebra in corresponding
theories which can be obtained as certain reductions of Hilbert space of the
$c = 1$  theories. From comparison with matrix models one could expect the
appearance of  $W_q$-algebra's for OPE's in different ($p,p'$)-minimal series
(see also [32]).

{\bf f)} Finally, it is interesting to study algebraic properties of the
Clebsh-Gordon coefficients for higher (at least, simply laced) groups. Namely,
do the $3j$-symbols
$C\left( ^{{\Bf \nu } _1{\Bf \nu } _2}_{{\Bf \xi } _1{\Bf \xi } _2}
\left| ^{{\Bf \nu } _3}_{{\Bf \xi } _3}\right.\right) $ with
additional conservation law like  ${\Bf \nu } _3 =
{\Bf \nu } _1 + {\Bf \nu } _2 - 2{\Bf \rho } $  or
${\Bf \nu } _3 = {\Bf \nu } _1 + {\Bf \nu } _2 - {\Bf \theta }$,  where
${\Bf \theta }$ -- the highest root of
$G$\footnote{Actually
the second choice allows corresponding operators to acquire
{\it unit} dimension instead of  $2{\Bf \rho } ^2$ for the first case, as it
has been
considered above, see also Appendix},
satisfy some clear algebraic relations like Jacobi identities\footnote{They
are antisymmetric for group $SU(n)$ only at  $n = 2, 5, 6, 9, 10,\ldots$
Thus, in these cases they have a chance to be a structure constants of Lie
algebra satisfying the Jacobi identities. At all remaining values of $n$
Clebsh-Gordon coefficients are symmetric and generate a commutative ring.}?

The answer to this question would state whether any chiral algebra can be
interpreted as universal envelope of some smaller and simpler algebraic
structure like $SU(2)$ case, when Clebsh-Gordon coefficients are the
structure constants (or adjoint representation) of the enveloping algebra
${\cal T}(\mu )$ of $SU(2)$ [1] (see also Appendix).

One could try also to guess, what can be this algebraic structure, just
generalizing the {\it result} of the KPW construction in the $SU(2)$
case. We add Appendix, describing immediate implication of such attempts:
concerning the Lie algebra, defined by symplectomorphisms of the coadjoint
orbits. This topic is not quite trivial and is of certain interest by its own,
though its relation to the KPW construction for  $G\neq SU(2)$  still remains
somewhat obscure.

\bigskip

We are indebted to A.Alekseev, Vl.Dotsenko, A.Gerasimov, D.Juriev, A.Losev,
S.Pa\-ku\-li\-ak and
A.Sagnotti for illuminating discussions.

\section*{Appendix. Symplectomorphisms as a {\it model}}
\def\theequation{A.\arabic{equation}}
\setcounter{equation}{0}

This Appendix is devoted to another ingredient of the KPW construction: the
interpretation of (the Lie-algebra fragment of) the algebra of observables as
the symplectomorphism algebra of a certain manifold. Symplectomorphisms
(area-preserving diffeomorphisms or Hamiltonian vector fields) usually form a
much wider space than more familiar isometries (since they preserve a
non-degenerate antisymmetric symplectic 2-form  $\omega _{ij}
dx^i\wedge dx^j$ rather than non-degenerate symmetric metric tensor
$g_{ij}dx^idx^j$). The symplectomorphisms algebra depends on the choice of
symplectic form  $\omega _{ij}$ but is usually infinite (as compared to finite
isometry groups). The symplectomorphisms are of special interest to the string
theory since they are natural generalizations of the Virasoro
algebra\footnote{Indeed, on complex manifolds the complex structure is
introduced whenever
a certain symplectic form is fixed (which satisfy additional relation
$\omega ^2 = -I)$ and, thus, the transformations which preserve the complex
structure are particular examples of symplectomorphisms.}.
In particular this is the reason why symplectomorphisms of  ${\Bf C}^2$ arise
in membrane theory {\it etc}.

The generators of the symplectomorphisms algebra can be represented either as
hamiltonians  $h(x)$  -- functions on a manifold or as hamiltonian vector
fields

\beq
e_h = \omega ^{ij}\partial _ih\partial _j\ .
\eeq
The Lie-algebra commutator is just a commutator of these vector fields

\beq
[e_{h_1},e_{h_2}] = e_{h_3} \footnote{Generically
there can be global obstacles for this relation.},
\eeq
or a Poisson bracket of the Hamiltonians

\beq
h_3 = \{h_1,h_2\} = \omega ^{ij}\partial _ih_1\partial _jh_2\ .
\eeq
Below we shall prefer the Poisson bracket representation.

We shall also restrict ourselves to the case of complex spaces  ${\Bf C}^n$
with coordinates  $z_1,\ldots,z_n,\bar z_1,\ldots,\bar z_n$ and
``flat" symplectic form
(or complex structure )  $\sum \ dz_i\wedge d\bar z_i$.

The study of this symplectomorphisms in the context of this paper is motivated
by the crucial discovery of [1,2] that the Lie-algebra piece of the algebra of
observables for  $G = SU(2)$  coincides with  $W_\infty $ represented as
symplectomorphism of  ${\Bf C}$. If it is translated to the algebraic language
this result implies that the algebra  ${\cal J}(M)$  of symplectomorphisms of
$M = {\Bf C}$  can be splitted (graduated) with respect to representations of
$G = SU(2)$  in such a way that any irreducible highest weight representation
appears exactly once, $i.e.$  ${\cal J}({\Bf C})$  is a representation of a
{\it model} of  $SU(2)$. This is a very nice statement of its own
interest and we are going to discuss and generalize it in this Appendix, namely
${\cal J}({\Bf C}^3)$  is closely related to a representation of a {\it model}
of $SU(3)$.

What distinguishes the KPW construction for the  $SU(2)$  case is
that the global  $SU(2)$  algebra is identified with the adjoint
representation from the $SU(2)$-{\it model}  ${\cal J}({\Bf C})$. In other
words the operators  $q_{1,m}$ would correspond to both  $O_{2{\Bf \rho } ,
{\Bf \xi } }$ and
$O_{{\Bf \theta },{\Bf \xi } }$ only in the  $SU(2)$ case (${\Bf \rho }  =
{1\over 2}\sum _{{\Bf \alpha} \in \Delta_{+}}{\Bf \alpha} $,  $\hbox{ }
{\Bf \theta}$ -- is the
highest positive root) when according to the first representation it has proper
conformal dimension while the second reflects its ``adjoint" properties. In
general  $2{\Bf \rho }  \neq  {\Bf \theta }$, therefore neither Toda fields
nor ghosts must
decouple for the operator  $O_{{\Bf \theta },{\Bf \xi } }$. However, this
identification
would still be valid if we use the construction which substitutes
$2{\Bf \rho } $  by
${\Bf \theta }$, but it does not seem to be quite adequate to the KPW
construction.
Usually, for  $G \neq  SU(2)$  the piece algebra of observables (consisting of
$O_{{\Bf \nu } ,{\Bf \xi } }$) is a  $G$-{\it model} with respect to external
action of global  $G$
(generated by  $\dst{Q_{\Bf \alpha}  = \int   {\cal J}_{\Bf \alpha} }$ and
$\dst{Q_{\Bf \nu }
= \int
{\cal H}_{\Bf \nu } }$), $i.e.$ it is not identifies with the OPE
involving the
operators  $O_{{\Bf \theta },{\Bf \xi } }$, which belong to the selected
set of observables
(we do not insist that this piece of observables form a
{\it Lie}-{\it algebra} !). However, despite all we have said, below in the
example for  $SU(3)$  we shall show that  $J({\Bf C}^3)$  is indeed a
kind of  $SU(3)$-{\it model} with respect to internal action of the adjoint
representation from  $J({\Bf C}^3)$. The relevance of these results to the KPW
construction is at least obscure, but we decided to present them here because
of their own interest.

Another remark is that it would be much more interesting to have a generic
theory of symplectomorphisms of the coadjoint orbits of  $G$  and to understand
what is the exact information about  $G$, encoded there. Our discussion of
${\Bf C}^N$ spaces below is just a first step in this direction.

Let us consider the space of Hamiltonians  $h(z_i,\bar z_i)$  on  ${\Bf C}^N$,
$i.e.$ just the space of all polynomials of  $z_i$ and  $\bar z_i$. Introduce
the gradation in this space by degrees of homogeneity in holomorphic and
anti-holomorphic coordinates. Consider now the degree two polynomials

\beq
h^{adj}_{ij} = z_i\bar z_j\ .
\eeq
Their Poisson brackets

\beq
\{h^{adj}_{ij},h^{adj}_{kl}\} = \delta _{il}h^{adj}_{kj} -
\delta _{jk}h^{adj}_{il}
\eeq
obviously describe the construction of  $GL(N)$  algebra and thus contain the
{\it adjoint} representation of  $SU(N)$  and the  $SU(N)$  scalar  $h_0 =
\sum |z_i|^2$. Associated vector fields

\beq
e^{adj}_{ij} = z_i\partial _j - \bar z_j\bar \partial _i
\eeq
obviously preserve our graduation, and the finite sets of polynomials  $h[m,n]$
form (probably reducible) representation of  $SU(N)$.

In order to select the irreducible representation we need to find out all
highest vectors, $i.e.$ the Hamiltonians annihilated by the action of  $e_{ij}$
with  $i<j$. Obviously the {\it holomorphic} functions of  $z_1$ and
$\bar z_N$ are among them

\beq
\left\{ h^{adj}_{ij},H(z_1,\bar z_N)\right\} = 0\ \ \ \hbox{for all }\  i<j\ .
\eeq
One can easily check that the highest vector

\beq
H_{pq} = z^p_1\bar z^q_N
\eeq
generates an irreducible representation with the highest weight

\beq
{\Bf \nu } _{pq} = p{\Bf \mu } _1 + q{\Bf \mu } _{N-1}\ ,
\eeq
where  $\{{\Bf \mu } _i\}$,  $i = 1,\ldots,N-1 = r = \hbox{rank} [SU(N)]$
are the fundamental
weights of  $SU(N)$. In particular  $H_{11} = z_1\bar z_N$ is the highest
vector of the adjoint representation with  ${\Bf \mu } _1 + {\Bf \mu } _{N-1}
 = {\Bf \theta }$.
Thus, we proved that  $J({\Bf C}^N)$  contains a piece of {\it model} of
$SU(N)$, consisting of all the representation of the form  (A9), with every
such representation arising {\it exactly once} (unless  ${\Bf \mu } _{N-1} \neq
{\Bf \mu } _1)$ This would be all irreducible representation for  $N=3$,
$i.e.$ for
$G = SU(3)$ (since all the representations are represented as  ${\Bf \nu }
=\sum ^r_{i=1}a_i{\Bf \mu } _i$ and for  $SU(3)$  $a_2,\ldots,a_{N-2}$ do not
need to vanish).

Several remarks are in order now. First,  $H(z_1,\bar z_N)$  do not exhaust the
entire set of Hamiltonians, annihilated by all  $e_{ij}$ with  $i<j$, the
simplest counterexample being the scalar  $h_0 = \sum |z_i|^2$. Therefore the
entire  $J({\Bf C}^N)$  is somewhat bigger that the {\it model} of
representations of
the form (A9): it can contain the extra scalars and thus extra representations
arising by multiplication with these scalars. So, a kind of factorization over
these scalars is required to make any  $SU(N)$  representation appearing only
once\footnote{Without
such factorization  $J({\Bf C}^N)$  seems to contain many copies
of the {\it model}.}.
Second, the above construction can be reduced by identification

\beq
z_1 = \bar z_N
\eeq
which does not break  $SU(N)$  commutation relations  (A5), but eliminates the
extra scalars. Note that this reduction does break our original gradation and
preserves only the degree of homogeneity in  $z_i$ and $\bar z_i$ together
rather than in  $z_i$ and  $\bar z_i$ separately. However, this reduction
eliminates also half of highest vectors:  $H(z_1,\bar z_N) \rightarrow  H(z_1)$
and we obtain a ``reduced" {\it model} of representations of the form

\beq
{\Bf \nu }  = p{\Bf \mu } _1
\eeq
In the particular case of  $SU(2)$  this is exactly what is necessary
to eliminate  (since  ${\Bf \mu } _1 = {\Bf \mu } _{N-1}$ for  $N = 2$).
Therefore the {\it model}
of  $SU(2)$  is just represented by

\beq
\hat J({\Bf C}) = J({\Bf C})/\{z_1 = \bar z_2)\ ,
\eeq
$i.e.$ by Hamiltonians of the form

\beq
h^{SU(2)}_{J,m} = z^{J+m}\bar z^{J-m}
\eeq
with (half-)integer  $J = \nu /\sqrt{2} \geq  0$  and  $m =
{\Bf \xi } /\sqrt{2}$,
$|m| \leq  J$. The adjoint representation is given by

\beq
h^{adj}_{1,m} = \{z^2\hbox{, } z\bar z\hbox{, } \bar z^2\}\ ,
\eeq
the highest vectors are

\beq
H_J \equiv  h_{J,J} = z^{2J}
\eeq
and thus  $\hat J({\Bf C})$  contains all the representations of (half-)integer
spin  $J$, $i.e.$ is the {\it model} of  $SU(2)$.

Third, as we already discussed, while in the case of  $G = SU(2)$  the above
algebra  $\hat J({\Bf C})$  is indeed the algebra of observables arising in the
KPW construction, it is not precisely  $J({\Bf C}^3)$  in the case  $G = SU(3)$
(despite that  $J({\Bf C}^3)$  is essentially the {\it model} of
$SU(3)$, note also that in contrast to the  $SU(2)$  case
$\hat J({\Bf C}^2) = J({\Bf C}^3)/\{z_1 = \bar z_3\}$  is {\it less} than a
{\it model} of  $SU(3)$ !). In particular, the commutator of some vectors
from the fundamental representation from  $J({\Bf C}^3)$  if defined by the
Poisson relations in  $J({\Bf C}^3)$  is vanishing, but this is far from being
obvious in the algebra of observables. For example, for  ${\Bf \nu }  =
{\Bf \mu } _1$,
${\Bf \nu } ' = {\Bf \mu } _2$ in general one has  ${\bf 3} \times  {\bf 3} =
{\bf 6} + \bar {\bf 3}$, and according
to the rule  ${\Bf \nu } '' = {\Bf \nu }  + {\Bf \nu }' - {\Bf \theta }$, only
{\bf 6}  is eliminated
by conservation law in the  $W_G$-Toda sector. Note, however, that this
particular discrepancy can be connected with a specific choice of symplectic
structure  $\omega _{ij} = \delta _{i \bar j}$  which does not include any
other invariant tensors, which should certainly
appear in the chiral KPW construction.


\begin{thebibliography}{99}
\bibitem{1} I.Klebanov, A.Polyakov  {\it Interaction of Discrete states in
Two-Dimensional String Theory},  Preprint PUPT-1281, September
1991

\bibitem{2} E.Witten  {\it Ground Ring of Two Dimensional String Theory},
Preprint IASSNS-HEP-91/51

\bibitem{3} R.Dijkgraaf, E.Verlinde, H.Verlinde  {\sl Comm.Math.Phys.}
{\bf 115} (1988) 649

\bibitem{4} P.Ginsparg  {\sl Nucl.Phys.} {\bf B295 [FS21]} (1988) 153

\bibitem{5} F.David  {\sl Mod.Phys.Lett.} {\bf A3} (1988) 1651

\bibitem{6} J.Distler, H.Kawai  {\sl Nucl.Phys.} {\bf B312} (1989) 509

\bibitem{7} B.Lian, G.Zuckerman  {\sl Phys.Lett.} {\bf B254} (1991) 417\\
 B.Lian, G.Zuckerman {\sl Phys.Lett.}  {\bf B266} (1991) 21\\
 P.Bouwknegt, J.McCarthy, K.Pilch  {\it Fock
Space resolutions of the Virasoro Highest Weight Modules with
$c\leq 1$},  Preprint CERN-TH-6196/91

\bibitem{8} E.Bergshoeff, C.Pope, L.Romans, E.Sezgin, X.Shen, K.Stelle
{\sl Phys.Lett.}
{\bf B243}  (1990) 350

\bibitem{9} G.Sotkov, M.Stanishkov  {\sl Nucl.Phys.} {\bf B356} (1991) 439\\
 G.Sotkov, M.Stanishkov,  C.J.Zhu  {\sl Nucl.Phys.} {\bf B356} (1991) 245

\bibitem{10} A.Gerasimov et al.  {\sl Nucl.Phys.} {\bf B360} (1991) 537

\bibitem{11} A.Bilal, V.Fock, I.Kogan  {\sl Nucl.Phys.} {\bf B359} (1991) 635

\bibitem{12} P.Mansfield, B.Spence  {\sl Nucl.Phys.} {\bf B362} (1991) 294

\bibitem{13} J.-L.Gervais, Y.Matsuo  Preprint LPTENS-91/35, NBI-HE-91-50

\bibitem{14} A.Morozov, M.Olshanetsky  {\sl Nucl.Phys.} {\bf B299} (1988) 389\\
 A.Gerasimov et al.  {\sl Int.J.Mod.Phys.} {\bf A5} (1990) 2495

\bibitem{15} A.Morozov, A.Rosly  {\sl Phys.Lett.} {\bf B214} (1988) 522

\bibitem{16} J.Cardy  {\sl Nucl.Phys.} {\bf B324} (1989) 581

\bibitem{17} A.Sagnotti  {\it Recent Developments in Open-String Theories},
Preprint ROM2F-91/11   and references therein

\bibitem{18} V.Kac  {\sl Lect.Notes in Physics} {\bf 94} (1979) 441

\bibitem{19} A.Rocha-Caridi  in: {\it Vertex Operators
in Mathematics and Physics},  {\sl MSRI
Publ.} {\bf 3}  (Springer, Heidelberg, 1984) 451

\bibitem{20} A.Morozov  {\sl Nucl.Phys.} {\bf B357} (1991) 619

\bibitem{21} D.Gepner, E.Witten  {\sl Nucl.Phys.} {\bf B278} (1986) 493

\bibitem{22} D.Zhelobenko  {\it Compact Lie Groups and Their Representations},
Moscow, Nauka,  1970

\bibitem{23} S.Lukyanov, V.Fateev  {\it Additional Symmetries in
Two-Dimensional
Conformal Field Theory and Exactly Solvable Models},  Preprints
ITP-88-74R,75R,76R, Kiev  1988

\bibitem{24} Vl.Dotsenko  {\sl Adv.Stud.in Pure Math.} {\bf 16} (1988) 123

\bibitem{25} V.Knizhnik, A.Polyakov, A.Zamolodchikov
{\sl Mod.Phys.Lett.} {\bf A3} (1988) 819

\bibitem{26} J.Thierry-Mieg  {\sl Phys.Lett.} {\bf B197} (1987) 368

\bibitem{27} P.Mansfield  {\sl Phys.Lett.} {\bf B248} (1990) 387

\bibitem{28} M.Bershadsky, H.Ooguri  {\sl Comm.Math.Phys.} {\bf 126} (1989) 49

\bibitem{29} A.Marshakov, A.Morozov  {\sl Nucl.Phys.} {\bf B339} (1990) 79

\bibitem{30} P.Forgacs, A.Wipf, J.Balog, L.Feher, L.O'Raifeartaigh
{\sl Phys.Lett.}
{\bf 227B} (1989)  214\\
J.Balog, L.Feher, L.O'Raifeartaigh, P.Forgacs, A.Wipf
{\sl Ann.Phys.} {\bf 203}  (1990) 76

\bibitem{31} V.Fateev, A.Zamolodchikov  {\sl Nucl.Phys.} {\bf B280} (1987) 644

\bibitem{32} D.Kutasov, E.Martinec, N.Seiberg
{\it Ground rings and their modules in $2d$
gravity with $c\leq 1$ matter},  Preprint PUPT-1293, RU-91-49, November
1991

\end{thebibliography}
\end{document}